\documentclass[iop]{emulateapj}

\usepackage{graphicx}
\usepackage{epstopdf}
\usepackage{rotating}

\shorttitle{Spectroscopy of High-Redshift Galaxies}
\shortauthors{Berry, Gawiser, etc.}

\begin{document}

\title{Stacked Rest-Frame UV Spectra of Ly$\alpha$-Emitting and Continuum-Selected Galaxies at $2<z<3.5$}

\author{Michael Berry\altaffilmark{\ref{Rutgers}},
Eric Gawiser\altaffilmark{\ref{Rutgers}}, 
Lucia Guaita\altaffilmark{\ref{Lucia}}, 
Nelson Padilla\altaffilmark{\ref{Catholica}}, 
Ezequiel Treister\altaffilmark{\ref{UH}}, 
Guillermo Blanc\altaffilmark{\ref{UT}}, 
Robin Ciardullo\altaffilmark{\ref{PSU}}, 
Harold Francke\altaffilmark{\ref{Catholica}}, 
Caryl Gronwall\altaffilmark{\ref{PSU},\ref{PSU2}}
}

\altaffiltext{1}{Physics and Astronomy Department, Rutgers University, Piscataway, NJ 08854-8019, U.S.A.
\label{Rutgers}}
\altaffiltext{2}{Oskar Klein Cosmology Centre, Department of Astronomy, Stockholm University, Stolkholm, Sweden
\label{Lucia}}
\altaffiltext{3}{Department of Astronomy and Astrophysics, Pontificia Universidad Catolica de Chile, Santiago, Chile
\label{Catholica}}
\altaffiltext{4}{Department of Astronomy, Universidad de Concepcion, Concepcion, Chile
\label{UH}}
\altaffiltext{5}{Department of Astronomy, University of Texas, Austin, TX 78712-0259, U.S.A.
\label{UT}}
\altaffiltext{6}{Deptartment of Astronomy and Astrophysics, The Pennsylvania State University, 525 Davey Laboratory, University Park, PA 16802, U.S.A.
\label{PSU}}
\altaffiltext{7}{Institute for Gravitation and the Cosmos, The Pennsylvania State University, University Park, PA 16802, U.S.A.
\label{PSU2}}

\begin{abstract}

We present properties of individual and composite rest-UV spectra of
continuum- and narrowband-selected star-forming galaxies (SFGs) at a
redshift of $2<z<3.5$ discovered by the MUSYC collaboration in the
Extended Chandra Deep Field-South. Among our sample of 81 UV-bright
SFGs, 59 have R$<$25.5, of which 32 have rest-frame equivalent widths
W$_{\mathrm {Ly\alpha}}>20{\mathrm \AA}$, the canonical limit to be
classified as a Ly$\alpha$-emitting galaxy (LAE). We divide our
dataset into subsamples based on properties we are able to measure for
each individual galaxy: Ly$\alpha$ equivalent width, rest-frame UV
colors, and redshift. Among our subsample of galaxies with R$<$25.5,
those with rest-frame W$_{\mathrm {Ly\alpha}}>20{\mathrm \AA}$ have
bluer UV continua, weaker low-ionization interstellar absorption
lines, weaker C IV absorption, and stronger Si II$^{\ast}$ nebular
emission than those with W$_{\mathrm {Ly\alpha}}<20{\mathrm \AA}$. We
measure a velocity offset of $\Delta v\sim600$ km s$^{-1}$ between
Ly$\alpha$ emission and low-ionization absorption, which does not vary
substantially among any of our subsamples. We find that the
interstellar component, as opposed to the stellar component, dominates
the high-ionization absorption line profiles. We find the low- and
high-ionization Si ionization states have similar kinematic
properties, yet the low-ionization absorption is correlated with
Ly$\alpha$ emission and the high-ionization absorption is not. These
trends are consistent with outflowing neutral gas being in the form of neutral clouds embedded in ionized gas as previously suggested by \cite{Steidel2010}. Moreover, our galaxies with bluer UV colors have stronger Ly$\alpha$ emission, weaker low-ionization absorption and more prominent nebular emission line profiles. From a redshift of $2.7<z<3.5$ to $2.0<z<2.7$, our subsample of galaxies with W$_{\mathrm {Ly\alpha}}<20{\mathrm \AA}$ show no significant evolution in their physical properties or the nature of their outflows. Among our dataset, UV-bright galaxies with W$_{Ly\alpha}>20{\mathrm \AA}$ exhibit weaker Ly$\alpha$ emission at lower redshifts, although we caution that this could be caused by spectroscopic confirmation of low Ly$\alpha$ equivalent width galaxies being harder at z$\sim$3 than z$\sim$2.

\end{abstract}

\section{Introduction}

Narrowband and broadband selection techniques around the Ly$\alpha$ emission line and Lyman limit respectively have made isolating Ly$\alpha$ emitters (LAEs) and UV-bright, continuum-selected galaxies quite effective at redshifts up to $z=7$. Through stacking photometric samples and stellar population modeling, numerous groups have uncovered a wealth of information on these galaxies' properties including stellar masses, star formation rates, ages, amount of dust ($E(B-V)$), and star formation histories \citep[e.g.,][]{Adelberger2000, Papovich2001, Shapley2001, Shapley2005, Erb2006, Reddy2006, Gawiser2006b, Pirzkal2007, Gawiser2007, Nilsson2007, Pentericci2007, Lai2008, Nilsson2009, Finkelstein2009, Guaita2010}. 
As a result, there have been numerous follow-up rest-frame ultraviolet (UV) spectroscopic surveys \citep[e.g.,][]{Shap2003, Erb2006, Kornei2010, Talia2011, Jones2011}. Rest-frame UV spectroscopy offers the ability to not only confirm redshifts, but also to study galactic-scale outflows, stellar photospheres and dust reddening through interstellar absorption line kinematics, the shape and strength of the Ly$\alpha$ emission line, P Cygni profiles from massive stars, and the UV spectral slope \citep{Leitherer1995, Pettini2000, Shap2003}. 

In order to determine the impact of feedback from star formation on galaxy evolution, we have targeted the peak of cosmic star formation density occuring at a redshift of $1.5<z<3$ when galaxies and their surrounding gas were evolving rapidly. Understanding the relation between star-forming galaxies and the surrounding intergalactic medium (IGM) at these redshifts will give insight into this relation at higher redshifts where spectroscopy is much more difficult. It is well known that outflows can occur in starburst galaxies in the local universe and, more recently, it has been shown that they are ubiquitous among high-redshift star-forming galaxies \citep{Steidel1996b, Pettini2000, Pettini2001, Shap2003, Martin2005, Rupke2005, Tremonti2007, Weiner2009, Steidel2010}. Analysis of these outflows has primarily been focused on modeling the complex spatial and velocity structures, the covering fraction, and the nature of neutral and ionized gas \citep[e.g.,][]{Heckman1990, Lehnert1996, Martin1999, Shap2003, Strickland2004, Schwartz2006, Grimes2009, Steidel2010}. Galactic-scale outflows are not only important in understanding how feedback from star formation affects galaxy evolution, but also how they affect the metal enrichment, ionization state, and physical state of the IGM.

In order to understand the connection between LAEs and continuum-selected galaxies, it is important to understand the different classifications for each type of galaxy, namely the criteria by which they are selected. Continuum-selected galaxies are chosen via color cuts and apparent magnitude limits, while LAEs are selected via equivalent width and line flux cuts that are independent of their colors. As defined by \cite{Steidel1996a}, Lyman break galaxies (LBGs) are continuum-selected galaxies at z$\sim$3 and typically have an apparent magnitude limit of R$<$25.5. \cite{Gawiser2006b} found LAEs at z$\sim$3 with L$_{\mathrm {Ly\alpha}}\ge 4\times10^{-17}$ ergs s$^{-1}$ cm$^{-2}$ to have a median apparent magnitude of $R=27$. In spite of this luminosity difference, \cite{Gawiser2006b} find that $z=3.1$ LAEs have similar rest-frame UV colors to LBGs, indicating that they are missed in continuum-selected surveys primarily due to their fainter continua. Similarly, \citep[][hereafter S03]{Shap2003} find that $20-25$ percent of their almost $1000$ LBGs have strong enough Ly$\alpha$ emission lines (W$_0 > 20{\mathrm \AA}$) to be classified as LAEs.

Given the faint nature of these galaxies (R$>23$), previous analyses have focused on stacking large numbers of low signal-to-noise spectra. Similar to stacking photometric samples, this technique yields the average properties of a 'typical' galaxy. S03 applied this technique to almost 1000 LBGs at a redshift of z$\sim$3 and found that they display a wide distribution of properties, with Ly$\alpha$ ranging from strong emission to absorption, varying UV continuum slopes, a range of velocity offsets between Ly$\alpha$ emission and interstellar absorption, and different strengths of interstellar absorption lines. Through dividing their sample into quartiles of Ly$\alpha$ equivalent width, they found that the sample with strongest Ly$\alpha$ emission had bluer UV continua, weaker low-ionization absorption, smaller kinematic offsets, stronger nebular emission, and older stellar populations.
With the exception of older stellar populations, these trends suggest that galaxies with stronger Ly$\alpha$ emission are less evolved, less rapidly forming stars, and likely less massive. Nonetheless, the presence of older stellar populations implies more evolution, indicating that galaxies with strong Ly$\alpha$ emission may be a heterogeneous population.

As the defining characteristic of LAEs, it is vital to understand the origin of the Ly$\alpha$ emission line, which is largely dependent on the effects of radiative transfer. Ly$\alpha$ photons are subject to resonant scattering and therefore have much longer pathlengths than continuum photons, making them more subject to extinction from dust. Thus, one explanation for LAEs' large Ly$\alpha$ equivalent widths is that they are young, chemically pristine galaxies undergoing one of their first major episodes of star formation. A second hypothesis suggests that they are older, more evolved galaxies where dust is segregated into clumps that are surrounded by a relatively dust-free intercloud medium \citep{Neufeld1991, Hansen2006, Finkelstein2009}. In this scenario, Ly$\alpha$ photons scatter off dust clumps while UV continuum photons penetrate them and are subject to extinction, thereby increasing the relative strength of the Ly$\alpha$ emission line. A lack of correlation between the spectral index of the UV continuum slope and Ly$\alpha$ equivalent width is interpreted as evidence for the decoupling of continuum and line extinction (S03) as would be expected if the neutral medium was inhomogeneous. For $2<z<4$ LAEs, \cite{Blanc2011} recently found E(B-V) to correlate with Ly$\alpha$ escape fraction in such a way that Ly$\alpha$ photons are being extinguished by the same amount as UV continuum photons, leading them to suggest that among LAEs, the Ly$\alpha$ equivalent width is neither being enhanced due to a clumpy ISM nor being preferentially quenched by dust. Finally, \cite{Malhotra2002} suggest that LAEs' large Ly$\alpha$ equivalent widths may instead be due to a peculiar initial mass function, zero-metallicity stars, or narrow-line active galactic nuclei.

Through stellar population modeling, LAEs have been shown to encompass a broad range of physical properties. At an epoch of $2<z<4.5$, many groups find that they are characterized as young ($<100$ Myr), relatively dust-free (E(B-V)$<0.3$) galaxies with low stellar masses ($<10^9~{\mathrm M}_{\odot}$) and low star formation rates ($2-3~{\mathrm M}_{\odot}$yr$^{-1}$) \citep{Gawiser2006b, Pirzkal2007, Gawiser2007, Nilsson2007}. Furthermore, LAEs have very high specific star formation rates ($7\times 10^{-9}$ yr$^{-1}$), defined as star formation rate per unit mass, as compared to LBGs ($3\times 10^{-9}$ yr$^{-1}$) indicating they are building up stellar mass at a faster rate than continuum-selected galaxies. This picture is consistent with LAEs being chemically pristine and at the beginning of an evolutionary sequence. On the other hand, various authors have found LAEs to exhibit a broader range of properties with ages ranging up to $>1$ Gyr, stellar masses up to $10^{10}~{\mathrm M}_{\odot}$, and A$_{1200}$ (defined as the absorption at 1200${\mathrm \AA}$) as high as 4.5 (\cite{Pentericci2007}; \cite{Nilsson2009}; \cite{Finkelstein2009}). These results indicate that some LAEs are older, dustier, more evolved galaxies similar to LBGs. Furthermore, \cite{Lai2008} and \cite{Guaita2010} found IRAC detected objects to exhibit properties similar to LBGs, motivating \cite{Lai2008} to suggest they may be a lower mass extension of LBGs.

In this paper, our goal is to understand the relationship among Ly$\alpha$ emission, interstellar kinematics, multiphase gas, nebular emission, and spectral slope for a sample of $2<z<3.5$ UV-bright star-forming galaxies. We subdivide our sample in terms of their Ly$\alpha$ equivalent width, rest-UV colors, and redshift to examine the relationships among these properties. This paper is organized as follows. In \S \ref{sec:methodology}, we present the spectroscopic sample including details of the observations and data reduction. We discuss the spectroscopic properties of our individual spectra in \S \ref{sec:ind_gal_prop}. In \S \ref{sec:stack} and \ref{sec:analysis}, we describe the stacking technique and analyze the composite spectra as subdivided based on Ly$\alpha$ equivalent width. Selection effects and trends among the outflowing component and galaxy properties are presented in \S \ref{sec:discussion}. Finally, in \S \ref{sec:conclusions}, we summarize our conclusions.

\section{Observations and Methodology}
\label{sec:methodology}
The galaxies that will be presented and analyzed in the following sections include narrowband-excess objects and continuum-selected sources. The latter category includes U band drop-outs at z$\sim$3 and galaxies with prominent Ly$\alpha$ decrements at z$\sim$2 \citep[LBGs and ``BX'' galaxies, continuum-selected UV-bright SFGs identified via the criteria of][]{Steidel1996a, Steidel2004}.
The imaging data used to select our targets were collected using the MOSAIC II CCD camera on the CTIO Blanco 4 m telescope (http://www.astro.yale.edu/MUSYC/). The narrowband data consisted of a series of exposures taken through $50{\mathrm \AA}$ wide full width at half-maximum (FWHM) filters at $\lambda 5007{\mathrm \AA}$ and $\lambda 4990{\mathrm \AA}$ for $z=3.1$ LAEs and $\lambda 3727{\mathrm \AA}$ for $z=2.07$ LAEs. The details of these observations are presented in \cite{Gronwall2007} and \cite{Ciardullo2011} for LAEs at $z=3.1$ and \cite{Guaita2010} at $z=2.07$. Our LBG selection criteria is presented in \cite{Gawiser2006a}, which covers a redshift interval of $2.6<z<3.5$, while our BX selection criteria is presented in \cite{Guaita2010} and spans $2<z<2.7$.
Our spectroscopic candidates were chosen from these two photometric surveys. In this paper, we present spectra taken from two spectroscopic surveys using the FORS and VIMOS instruments on the Very Large Telescope (VLT). The first survey, using the FORS instrument, targeted primarily z$\sim$2 LAEs and BXs, along with a few z$\sim$3 LAEs and LBGs. The primary targets for observation using the VIMOS instrument were X-ray sources analyzed in \cite{Treister2009} with z$\sim$3 LAE and LBG candidates used to fill the remaining slitlets.

\subsection{FORS Spectroscopy}
\label{sec:FORS}
For our FORS spectroscopic sample, the primary targets were chosen by first prioritizing $z=2.1$ LAE candidates, followed by BX candidates and finally z$\sim$3 LAEs and LBGs were used to fill available slitlets. The targets covered six $7'\times 7'$ masks in the CDF-S, however one of these masks was unusable due to poor observing conditions and has been omitted from the dataset. Each of the five remaining slitmasks consisted of $\sim30$ targets with $\sim70\%$ expected to be at z$\sim$2.

These data were obtained using the FORS2 spectrograph on UT1 with the B1200 grism, the e2V UV-sensitive CCD, an atmospheric dispersion corrector, and $1.''0$ slitlets. This configuration led to a dispersion of $0.86{\mathrm \AA}$ per pixel corresponding to a nominal spectral resolution of $3.4{\mathrm \AA}$ (FWHM) or R$\sim$1400 over a spectral coverage of $3200 - 6000 {\mathrm \AA}$ for each slitlet. The small angular size of our targets caused the actual spectral resolution to be seeing limited in almost all cases, which varied from $0.''7$ to $1.''2$ leading to a minimum spectral resolution of 2.4${\mathrm \AA}$. The observations were taken in a series of four exposures with exposure times of $1800-2400$ seconds per mask. We also note that the throughput at $\lambda 3727{\mathrm \AA}$ is $\sim$2 times worse than that at $\lambda 5000 {\mathrm \AA}$.

The data were reduced using the standard FORS pipeline. This process included bias-subtraction, flat fielding of science data, sky subtraction, and wavelength calibration through arc lamp exposures. For 5 objects, we had difficulty wavelength calibrating the $\lambda 3200 - 4200{\mathrm \AA}$ region. We found an optimal wavelength calibration by lowering the threshold for identifying arc lamp lines and adding two lines to the arc lamp file at $3612.87{\mathrm \AA}$ (Cd I) and $3663.00{\mathrm \AA}$ (Hg I). We removed three objects that had data in this region for which we were not able to correctly calibrate their wavelengths. The spectra were then divided into individual spectrograms whereby one dimensional spectra were extracted using IRAF's apall task. Finally, each one-dimensional spectrum was corrected for atmospheric extinction and flux calibrated using observations of spectrophotometric standards observed with a $1.''5$ long slit using the same B1200 grating. A one sigma error array was then extracted according to the FORS reduction pipeline.

\subsection{VIMOS Spectroscopy}
\label{sec:VIMOS}
Targets observed with the VIMOS instrument were distributed across four slitmasks in the CDF-S, each with a field of view of $7' \times 8'$, covering an area of $15' \times 16'$. These observations typically targeted $\sim$5 LAEs and $\sim$40 LBGs per mask for a total of 23 LAEs and 164 LBGs. We used the VIMOS spectrograph with the MR grism and an OS-blue filter mounted at the Nasmyth focus B of UT3 to obtain these spectra. The pixel scale of the EEV 4k x 2k CCD was $0.''2$ pixel$^{-1}$, which, in conjunction with the MR grism, led to a dispersion of $2.57{\mathrm \AA}$ per pixel (R$\sim400$) and a spectral coverage of $4300 - 6800 {\mathrm \AA}$ on each slitlet. The nominal spectral resolution in combination with the $1.''0$ slits is $12.5{\mathrm \AA}$ (FWHM). 
For these observations, the seeing was required to be better than $1''$. Each mask was observed for three hours and eighteen minutes. The VIMOS data were reduced using the standard ESO VIMOS pipeline version 2.1.6. For more information on the observations and reduction, see \cite{Treister2009}. One dimensional spectra were then extracted using IRAF's apall task.

The VIMOS spectra were not flux calibrated to a spectrophotometric standard. As a result, we empirically flux calibrated the individual spectra based upon the FORS spectra. 
This procedure consisted of fitting a powerlaw to the $\lambda > 1220{\mathrm \AA}$ rest-frame continuum of FORS and VIMOS spectra via a least-squares algorithm. The individual VIMOS galaxy spectral slopes were then adjusted by the difference between the median FORS and VIMOS spectral slopes ensuring the same relative spectral slope. 
The VIMOS spectra were then normalized individually by convolving their fluxes (F$_{\lambda}(\lambda)$) with the V-band transmission curve, then rescaled to match the observed V-band magnitude.

Due to the artificial flux calibration, we also extracted the one sigma error spectrum empirically. For each spectrum, a $515{\mathrm \AA}$ wide top hat, corresponding to $200$ pixels or $40$ resolution elements, was shifted down the spectrum wherein, for each pixel, we recorded the median number of counts and relative root mean square deviation. Our error spectrum is conservative as we only masked the Ly$\alpha$ emission line and the $\lambda 5577$ skyline, but no other features. To check the accuracy of empirically extracting the error spectrum, we applied the same routine covering the same number of resolution elements to the FORS spectra. This process recovered an error spectrum comparable to the formal error spectrum produced by the FORS reduction pipeline.

\subsection{Redshift Identification}
\label{sec:redshift_id}
Spectral identification and redshifts were assigned by interactively examining the observed spectra in both one and two dimensions. All redshifts were independently confirmed by two investigators. Because our spectra lack systemic redshifts from nebular emission lines, redshifts were determined from the Ly$\alpha$ emission line. There were four continuum-selected galaxies with interstellar absorption lines and no Ly$\alpha$ emission. For these objects, redshifts were determined from the observed wavelength of the Si II $1260{\mathrm \AA}$ and C II $1334{\mathrm \AA}$ absorption lines, as they are the strongest lines that are not affected by blending. We then calculate their expected Ly$\alpha$ emission line redshift by applying a median offset of Ly$\alpha$ emission and interstellar absorption ($<v> = 650$ km s$^{-1}$) found to be characteristic of z$\sim$3 LBGs by S03. For stacking purposes, we use each object's Ly$\alpha$ emission line redshift to shift them into the rest-frame. We note that varying kinematic offsets between emission and absorption will artificially broaden the absorption lines.

For a large fraction of our LAEs, redshifts are determined solely from the Ly$\alpha$ emission line. There are many dangers involved in single line redshifts, which makes it very important to correctly identify this line as high-redshift Ly$\alpha$. Since we use an [O II] $\lambda 3727$ and an [O III] $\lambda 5007$ filter to select redshift z=2.1 and 3.1 LAEs, it is unlikely that local [O II] and [O III] emitting galaxies will survive the color cuts. However, intervening high-equivalent width [O II] $\lambda 3727$ emission line galaxies ($z=0.34$) can be mistaken as z=3.1 LAEs. The [O II] doublet $\lambda \lambda$ 3726, 3729 is just resolved in our FORS $z=3.1$ sample, which would identify the galaxy as a low-redshift [O II] emitting interloper. Unfortunately, the VIMOS spectral resolution is insufficient to resolve the two components of the doublet, making it difficult to distinguish [O II] from Ly$\alpha$. \cite{Gawiser2006a} and \cite{Gronwall2007} do not find any interlopers among LAEs observed with the Inamori Magellan Areal Camera and Spectrograph (IMACS) covering a wavelength range $4000-10000{\mathrm \AA}$ that would show multiple emission lines. Nonetheless, as the transmission fraction of the IGM due to H I absorption is $\sim$0.7 at $z=3$ \citep{Madau1995}, we can identify [O II] $\lambda 3727$ emitting galaxies from their lack of a Lyman decrement. We do not find any low-redshift emission line interlopers. Among the targets observed with the FORS spectra, we spectroscopically confirmed 13 of 36 $z=2.1$ LAEs, 24 of 55 BXs, 23 of 25 $z=3.1$ LAEs and 3 of 16 LBGs. Of the z$\sim$3 VIMOS targets, we spectroscopically confirmed 17 of 23 LAEs and 40 of 132 LBGs.

\section{Individual Galaxy Properties}
\label{sec:ind_gal_prop}

In this section, we analyze the individual galaxy spectra included in the composites. For individual spectra, the faint nature of high-redshift SFGs limits us to studying trends based on only a few parameters that are measurable for each galaxy: Ly$\alpha$ equivalent width, redshift, rest-frame UV color, and rest-UV continuum luminosity. For certain galaxies with Ly$\alpha$ emission and interstellar absorption, we are also able to measure a relative velocity offset. Due to the complexities associated with selection effects, we do not attempt to study the relationship between rest-UV continuum luminosity and other galaxy properties. As a maximum apparent magnitude of R$<$25.5 is used for broadband-selected objects, the rest-UV luminosity threshold for z$\sim$2 BX galaxies will be $\sim$0.5 magnitudes brighter than for z$\sim$3 LBGs. Additionally, we find that 5 BX galaxies, 3 $z=2.1$ LAEs, and 4 LBGs, i.e., $22\%$ of our R$<$25.5, Ly$\alpha$ emission line galaxies, show multiple-peaked Ly$\alpha$ emission. Similarly, \cite{Kulas2011} found that $\sim30\%$ of their Ly$\alpha$ emission line galaxies at $2<z<3$ showed multi-peaked Ly$\alpha$ emission. We also note seeing evidence for fine-structure Si II$^{\ast}$ emission in the individual spectra of 7 BX galaxies and 3 LBGs.

\begin{figure}[!t]
  \plotone{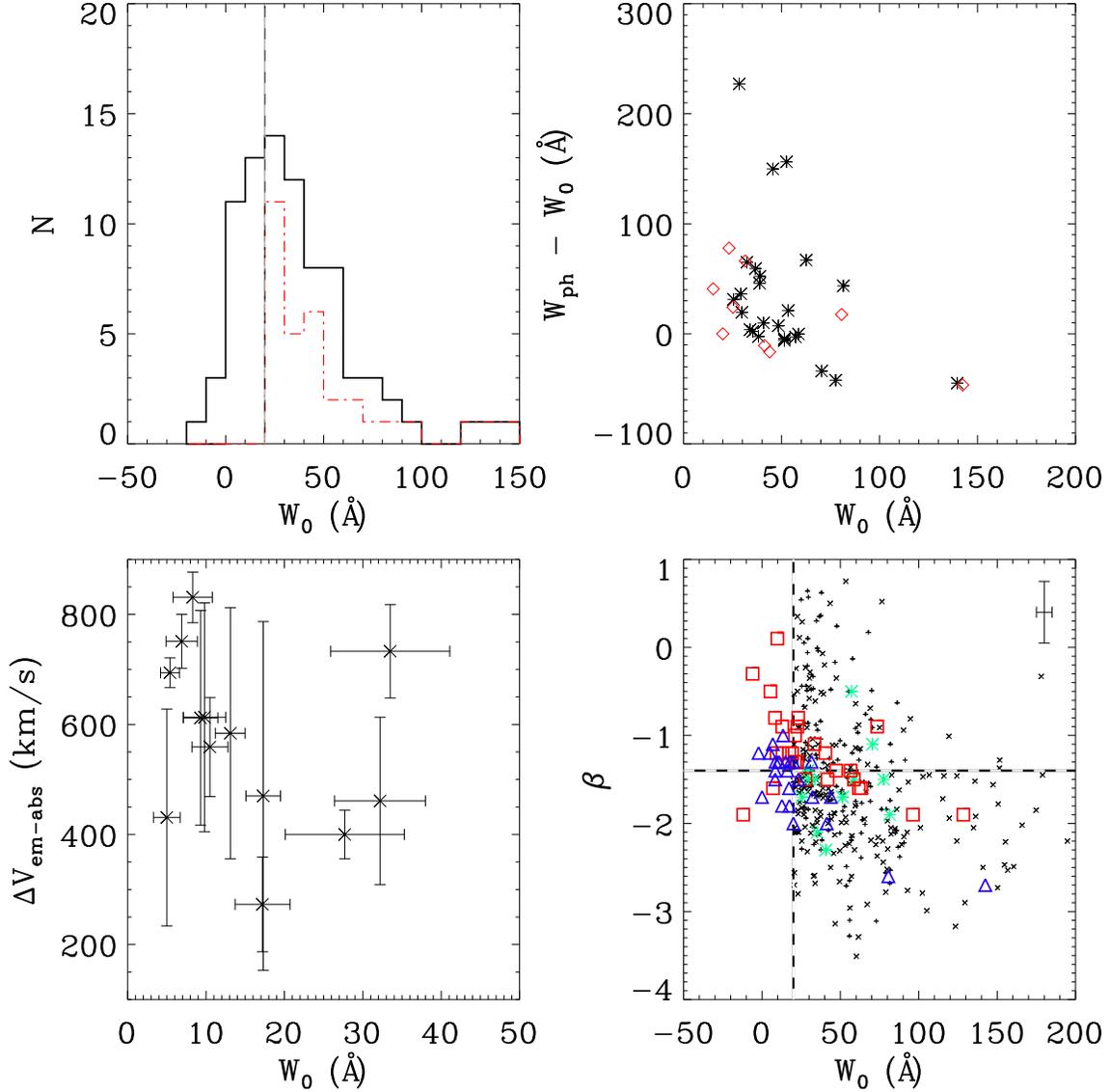}
    \caption{Top left: the rest-frame Ly$\alpha$ equivalent width ($W_0$) distribution for our spectroscopic sample (black solid line) and our UV-bright LAE subsample (red dot-dashed line). All reported equivalent widths are in ${\mathrm \AA}$. Top right: $W_{{\mathrm phot}}$ v. $W_0$ for z$\sim$3 LAEs (black stars) and z$\sim$2 LAEs (red diamonds). Bottom left: $\Delta v_{{\mathrm {em-abs}}}$ v. spectroscopic $W_0$ where errors in $\Delta v_{{\mathrm {em-abs}}}$ are the range values found for different transitions. Bottom right: spectral slope index, $\beta$ (F$_{\lambda}~\propto~\lambda^{\beta}$), as calculated from the photometric colors, v. W$_0$ for R$>$25.5 galaxies (cyan crosses), z$\sim$3 R$<$25.5 galaxies (red squares), and z$\sim$2 R$<$25.5 galaxies (blue triangles). A typical error measurement is plotted in the upper right corner. LAEs from \cite{Guaita2011} (black X's) and \cite{Nilsson2009} (black crosses) are also plotted. The vertical line is the traditional cut for LAEs (W$_0=20{\mathrm \AA}$) and the horizontal line is the division between the blue and red spectral slope composites.}
\label{Fig:ind_gal_prop}
\end{figure}

In 8 of 17 BX galaxies and 15 of 28 LBGs, the only visible feature is Ly$\alpha$ emission. In 2 BXs and 2 LBGs, we observe multiple absorption lines and no Ly$\alpha$ emission, and in 3 of these 4 galaxies, Ly$\alpha$ appears as broad absorption. In these cases, interstellar absorption lines are used to measure the redshift. In the remaining 7 BXs and 11 LBGs, Ly$\alpha$ emission and interstellar absorption are present allowing a measurement of the relative velocity offset. For 77 of 81 galaxies, redshifts were measured from the Ly$\alpha$ emission line. This fraction differs significantly from S03, as nearly one third of their LBG spectroscopic sample had Ly$\alpha$ only in absorption. This difference is likely due to lower S/N in our spectra making absorption line-only redshifts difficult. The top left panel of Figure \ref{Fig:ind_gal_prop} shows the distribution of Ly$\alpha$ equivalent widths for our entire spectroscopic sample (black, solid line) and UV-bright LAEs (red, dot-dashed line). We discuss the impacts of the different Ly$\alpha$ equivalent width distribution in \S \ref{sec:ind_lya_prop}.

Among the narrowband-selected galaxies, in all except for one, the only visible spectroscopic feature is Ly$\alpha$ emission. This is likely due to the intrinsically faint nature of their continua. In the top right panel of Figure \ref{Fig:ind_gal_prop}, photometric versus spectroscopic rest-frame equivalent width is plotted for z$\sim$3 LAEs as black stars and z$\sim$2 LAEs as red diamonds. We find that for z$\sim$3 LAEs, photometric equivalent widths are systematically overestimated as the median photometric equivalent width is $W_{\mathrm {phot,3}}=69.2{\mathrm \AA}$ while the median spectroscopic equivalent width is $W_{\mathrm {spec,3}}=36.6{\mathrm \AA}$. However, for z$\sim$2 LAEs, these two equivalent width measurements are consistent as $W_{\mathrm {phot,2}}=27.3{\mathrm \AA}$ and $W_{\mathrm {spec,2}}=31.5{\mathrm \AA}$. Photometric equivalent widths are calculated by using broadband filters (U+B at z$\sim$2, and B+V at z$\sim$3) to estimate the continuum flux density and a narrowband filter to estimate the amount of excess Ly$\alpha$ flux. For details on the photometric equivalent width calculations, see \cite{Gronwall2007, Guaita2010}. For particularly faint objects, spectroscopic equivalent widths may be biased low as a result of fake continuum in individual spectra. Since we weight each object in the composites by signal divided by noise squared, these objects will have a minimal effect on the composite spectra of which all equivalent widths are measured spectroscopically. A more detailed analysis of photometric and spectroscopic equivalent widths is necessary to fully understand the cause of this discrepancy.

In galaxy spectra with both Ly$\alpha$ emission and interstellar absorption, we measure the average relative velocity offset, $\Delta v_{{\mathrm {em-abs}}}$. In the bottom left panel of Figure \ref{Fig:ind_gal_prop}, the distribution of $\Delta v_{{\mathrm {em-abs}}}$ as a function of Ly$\alpha$ equivalent width is plotted where the errors represent the range in velocity offsets measured from different transitions. The median velocity offset is $\Delta v_{{\mathrm {em-abs}}}= 584$ km s$^{-1}$, similar to S03 who found $\Delta v_{{\mathrm {em-abs}}}= 650$ km s$^{-1}$. In all of our galaxies, Ly$\alpha$ emission is at a higher redshift than interstellar absorption. This result is consistent with S03 who find that nearly all of their 323 LBGs show Ly$\alpha$ at a higher redshift than interstellar absorption. These velocity offsets indicate that most likely the emission is coming from the farside component of the outflow while interstellar absorption, Ly$\alpha$ absorption, and blueshifted Ly$\alpha$ emission are coming from the nearside component. Large-scale outflows are caused by feedback from star formation and supernovae, as first reported in z$\sim$3 LBGs by \cite{Steidel1996b}. In general, stellar systemic redshifts are measured from photospheric features from hot stars. In the rest-UV, photospheric features are too weak to be seen in individual spectra. Systemic redshifts can also be determined from rest-optical nebular emission lines, but these were not available. Therefore, in the rest of our analysis, we measure the relative velocity offset between Ly$\alpha$ emission and absorption. While velocity offset tends to decrease with increasing Ly$\alpha$ equivalent width, we do not find this trend to be very significant. Furthermore, none of the UV-bright SFGs with $W_{\mathrm {Ly\alpha}}>40{\mathrm \AA}$ have visible interstellar absorption lines.

The spectral slope of the UV continuum is largely dependent on the star-formation history and amount of extinction. For UV-bright, star-forming galaxies, the unreddened UV continuum remains fairly constant for ages 10 Myr to 1 Gyr where the spectral slope index, $\beta$ (F$_\lambda~\propto~\lambda^\beta$), ranges from -2.6 to -2.1 \citep[based on models of SFGs from][]{Leitherer1999}. Both narrowband- and continuum-selected galaxies at $2<z<3.5$ have typical ages within this range \citep{Papovich2001, Shapley2001, Erb2006, Gawiser2007, Kornei2010} indicating that large differences in UV spectral slope reflect varying amounts of dust extintion.
For a large number of our galaxies, the individual spectra are of insufficient S/N to accurately measure a UV continuum slope. However, using broadband photometry, BVRI for z$\sim$3 and UBVRI for z$\sim$2 SFGs, we calculate the UV spectral slope index after correcting for IGM extinction and Ly$\alpha$ equivalent width for each galaxy. The bottom right panel of Figure \ref{Fig:ind_gal_prop} shows the distribution of UV spectral slopes as a function of Ly$\alpha$ equivalent width for R$<$25.5, $2<z<2.7$ galaxies (blue triangles), R$<$25.5, $2.7<z<3.5$ galaxies (red squares), and R$>25.5$ galaxies (cyan crosses) with a typical error measurement in the upper right corner. LAEs from \cite{Guaita2011} (black X's) and \cite{Nilsson2009} (black crosses) are also plotted. We find a median spectral slope of $\beta = -1.4$, then create a red ($\beta > -1.4$) and a blue ($\beta \le -1.4$) spectral slope subsample. Galaxies with nondetections in the B, V, and/or R bands have been omitted from this plot. We find a large range in $\beta$ values from $\beta=-2.7$ to $\beta=0.1$, however, for galaxies with rest-frame W$_{\mathrm {Ly\alpha}}>20{\mathrm \AA}$, $\beta$ ranges from $\beta=-2.7$ to $\beta=-0.8$ indicating a relation between stronger Ly$\alpha$ emission and bluer spectral slopes. The minimum observed spectral slopes are in good agreement with model predictions for a dust-free SFG with an age of 10 Myr \citep[$\beta \sim -2.6$,][]{Leitherer1999}. UV-bright SFGs with W$_{\mathrm {Ly\alpha}}<20{\mathrm \AA}$ subset have a mean spectral slope of $<\beta>=-1.22\pm0.47$ while UV-bright SFGs W$_{\mathrm {Ly\alpha}}>20{\mathrm \AA}$ have a bluer mean spectral slope of $<\beta>=-1.52\pm0.45$ where errors represent the standard deviation on the mean. Moreover, the red and blue spectral slope subsamples have median Ly$\alpha$ equivalent widths of W$_0=15.1{\mathrm \AA}$ and W$_0=32.2{\mathrm \AA}$ respectively. These trends display a relation between increasing Ly$\alpha$ equivalent width and bluer spectral slope, which has previously been observed in z$\sim$3 LBGs \citep{Shap2003, Steidel2010, Kornei2010}.

Through dividing the sample into two redshift bins, $2.0<z<2.7$ and $2.7<z<3.5$, we find that galaxies at both redshifts show a similar relationship between Ly$\alpha$ equivalent width and spectral slope. Surprisingly, galaxies at z$\sim$2 show bluer continuum slopes on average ($<\beta> = -1.87\pm0.51$) than galaxies at z$\sim$3 ($<\beta> = -1.34\pm0.33$), which appears as a systematic offset independent of Ly$\alpha$ equivalent width and is further discussed in \S~\ref{sec:red_evol}. This trend is still observed when using the same observed photometry (BVRI) to determine rest-UV slopes at both redshifts.

\section{Stacking Procedure}
\label{sec:stack}
In generating the following composite spectra, we begin by identifying a subsample of galaxy spectra to be combined. We exclude LAEs whose spectra are plagued by poor sky subtraction or have minimal wavelength coverage. Of the 52 spectroscopically confirmed LAEs, we include 26 observed with the FORS instrument and 10 with the VIMOS instrument. Additionally, SFGs lacking precise redshifts determined by either the Ly$\alpha$ emission line or several interstellar absorption lines are also excluded from the composite. These criteria yield 28 VIMOS LBGs and 17 FORS BXs out of 73 spectroscopically confirmed continuum-selected galaxies. The effects of omitting LBGs without detectable Ly$\alpha$ emission redshifts is discussed in section \ref{sec:sel_eff}.

We proceed to shift the extracted, one-dimensional, flux-calibrated spectra into the rest frame. The spectra are subsequently rebinned to a dispersion of $1{\mathrm \AA}$ per pixel corresponding to roughly the FORS resolution. We then renormalize them to their $1250-1300{\mathrm \AA}$ continuum. Outlying data points due to cosmic-ray events and sky subtraction residuals are systematically masked, totaling less than 0.2$\%$ of the data. Finally, using an optimal weighting routine where weights are defined as $1250-1300{\mathrm \AA}$ continuum strength divided by noise squared, we co-add the individual spectra \citep{Gawiser2006a}. We note that a variety of UV spectral slopes in conjunction with the majority of the VIMOS spectra covering only $\lambda<1600 {\mathrm \AA}$ causes the S/N to drop rapidly and the continuum level to become uncertain above this wavelength.

In order to understand the connection between UV-bright star-forming galaxies with and without strong Ly$\alpha$ emission, we divided our dataset into three subsamples: UV-bright non-LAE, UV-bright LAE, and UV-faint LAE. In Table \ref{Tab:sel_crit} we present our selection criteria for and the number of galaxies in each composite. We also define a UV-bright SFG subsample with R$\le$25.5 which is composed of the UV-bright non-LAE and UV-bright LAE subsamples. The UV-bright non-LAE and UV-bright LAE subsets differ slightly in magnitude, $<R_{UV-bright non-LAE}>=24.46$ and $<R_{UV-bright LAE}> =24.84$ with a scatter of $\pm0.68$ and $\pm0.42$ respectively, although this difference is not significant. This difference is in part due to the fact that larger Ly$\alpha$ line fluxes make spectroscopic identification easier. Selection effects based on luminosity and Ly$\alpha$ equivalent widths are discussed in detail in \S \ref{sec:sel_eff}. The primary goal for these different spectroscopic subsamples is to understand the effects of increasing Ly$\alpha$ emission and decreasing magnitude on galaxy properties. We do note that our R$\le 25.5$ LAE subsample contains galaxies with a broad range of properties including galaxies with strong interstellar absorption lines, Si II$^{\ast}$ fine-structure emission, and faint continua where the only feature seen is Ly$\alpha$ emission. We also subdivide our dataset into a red ($\beta > -1.4$) and blue ($\beta \le -1.4$) spectral slope subsample based on their intrinsic rest-UV colors. We exclude 6 narrowband-selected galaxies with B, V, or R band magnitudes greater than 1 magnitude from these composites as their spectral slopes are not well constrained.

\section{Analysis of Composite Spectra}
\label{sec:analysis}

The absorption and emission features observed in these spectra are produced in gas associated with galactic-scale outflows, H II regions, and the winds and photospheres of massive stars. Figure \ref{Fig:uber_stack} shows the UV-bright SFG composite spectrum of 59 R$\le$25.5 SFGs. The most prominent features in the rest-UV composite spectra are blueshifted interstellar absorption and H I Ly$\alpha$. These features trace large-scale outflows of interstellar medium, which have been modelled in detail for high-redshift LAEs and LBGs \citep{Verhamme2008, Laursen2011}. In individual galaxy spectra, due to low S/N, only redshifts can be determined from strong interstellar absorption. However, with higher S/N, composite spectra allow for more robust measurements of their equivalent widths as well as revealing fainter features such as nebular emission.

In Figure \ref{Fig:comp3_stack}, the top panel displays our composite UV-bright non-LAE spectrum (27 UV-bright non-LAEs), the middle panel shows our UV-bright LAE composite spectrum (32 UV-bright LAEs) and the bottom panel shows our UV-faint LAE composite spectrum (22 LAEs). For our UV-faint LAE composite spectrum, limited spectral coverage in conjunction with all of these LAEs being at $z=3.1$ causes our sample size to drop to 8 galaxies ($35\%$) at $\lambda <1340 {\mathrm \AA}$. In addition, we have masked the region around $1360 {\mathrm \AA}$ due to residuals from the $\lambda 5577 {\mathrm \AA}$ skyline. For this reason, we limit our analysis of the LAE composite to wavelengths blueward of $1340 {\mathrm \AA}$.

Across the UV-bright LAE and UV-bright non-LAE composites, the most striking trends are the decrease in interstellar absorption and increase in Si II$^{\ast}$ nebular emission with increasing Ly$\alpha$ emission. In this section, we first focus our analysis on the Ly$\alpha$ emission and absorption line profile. We then discuss low- and high-ionization interstellar absorption line features. Finally, we proceed to discuss nebular features and fine-structure Si II$^{\ast}$ emission within the systemic component of the galaxies.

        \subsection{Lyman Alpha Emission and Absorption}
\label{sec:lya}

Ly$\alpha$ photons originate from recombinations in H II regions around massive stars, yet a number of effects impact the strength, shape, and relative velocity of the emergent Ly$\alpha$ profile; including intrinsic star formation rate, amount of dust, dust geometry, covering fraction, and the kinematics of the outflowing gas. 
For a more in depth discussion on the shape of the Ly$\alpha$ profiles in this dataset, see Gawiser et al. (2012) [in preparation]. Nonetheless, we are still equiped with the overall strength of Ly$\alpha$ emission and absorption as well as its relative velocity offset with respect to other spectroscopic features.

\begin{figure}[!t]
  \plotone{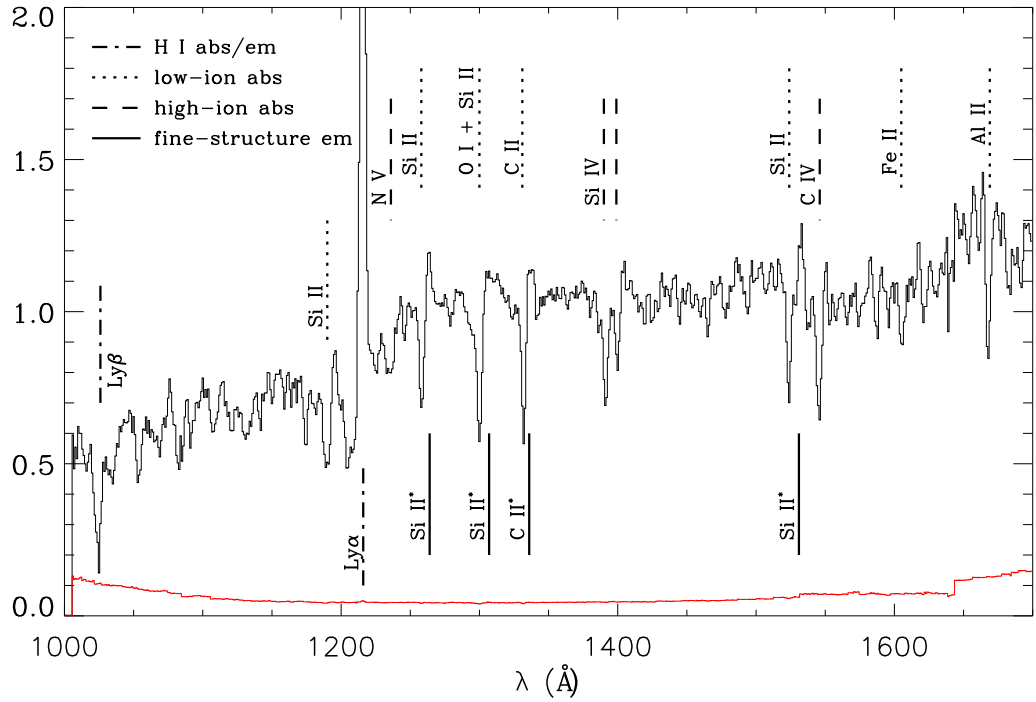}
    \caption{Composite continuum-normalized rest-frame UV spectrum of 59 SFGs (R$<$25.5) with the composite one sigma error spectrum plotted in red. H I Ly$\alpha$ emission and low- and high-ionization interstellar absorption originating from the outflowing component are the most prominent features. Weaker features such as stellar emission from P Cygni profiles and fine-structure emission (Si II$^{\ast}$ and C II$^{\ast}$) are also visible.}
\label{Fig:uber_stack}
\end{figure}

For our UV-bright SFG composite (Fig.~\ref{Fig:uber_stack}), we measure a rest-frame Ly$\alpha$ emission equivalent width of W$_o = 19.1\pm1.5 {\mathrm \AA}$. We also note that there is evidence for weak blueshifted emission at a velocity offset of $\Delta v_{\mathrm {em-abs}}\sim5000$ km s$^{-1}$ separated by Ly$\alpha$ absorption. Ly$\alpha$ photons are much more likely to escape from a galaxy if their relative velocities have been shifted off resonance with respect to the bulk component of the neutral material. We find it interesting that there is blueshifted emission at a very high velocity offset as the UV-bright SFG composite represents an average of all the UV-bright galaxies including several with strong Ly$\alpha$ absorption. For the main Ly$\alpha$ absorption, we measure an equivalent width of W$_o = -2.7\pm0.3 {\mathrm \AA}$ at a velocity offset of $\Delta v_{\mathrm {em-abs}}\sim2400$ km s$^{-1}$. Equivalent widths and velocity offsets for our UV-bright SFG composite spectra are presented in Table \ref{Tab:all_SFG_prop}. Since the composite spectra include galaxies observed with the FORS and VIMOS instruments, we smooth all spectra by the VIMOS resolution ($v\sim650$ km s$^{-1}$), which we calculate to be 3.0${\mathrm \AA}$ under optimal observing conditions. We also set an upper bound of 3.1${\mathrm \AA}$ on our spectral resolution as the minimum FWHM value measured among the strong interstellar features. Assuming an effective spectral resolution of 3.0${\mathrm \AA}$ and subtracting the instrument FWHM from the observed FWHM in quadrature, we compute a deconvolved FWHM(Ly$\alpha$) = $530\pm30$ km s$^{-1}$ in the UV-bright SFG composite. While the UV-bright SFG composite represents the average strength of Ly$\alpha$ emission among our R$<$25.5 SFGs, this is not representative of the entire population of SFGs due to selection effects. For instance, spectroscopic identification is easier using an emission line rather than an absorption line. As objects become fainter, the fraction of galaxies with Ly$\alpha$ in emission increases. We discuss the implications of this effect in \S \ref{sec:sel_eff}. Additionally, the weight causes brighter galaxies to contribute more to the composite. These factors have opposing effects on the strength of the Ly$\alpha$ feature in Figure \ref{Fig:uber_stack}, so it will not necessarily be that of the average SFG. However, we do find the median Ly$\alpha$ equivalent width of the UV-bright SFG sample (W$_0=21.0{\mathrm \AA}$) to be consistent with that of the UV-bright SFG composite. A wide distribution of Ly$\alpha$ equivalent widths and line profiles is seen in our individual spectra, which we discuss in section \ref{sec:ind_lya_prop}. In \S \ref{sec:Lya_dep}, we discuss the relationships among the strength of the Ly$\alpha$ line and other galaxy properties.

\begin{figure}[!t]
\epsscale{0.8}
  \plotone{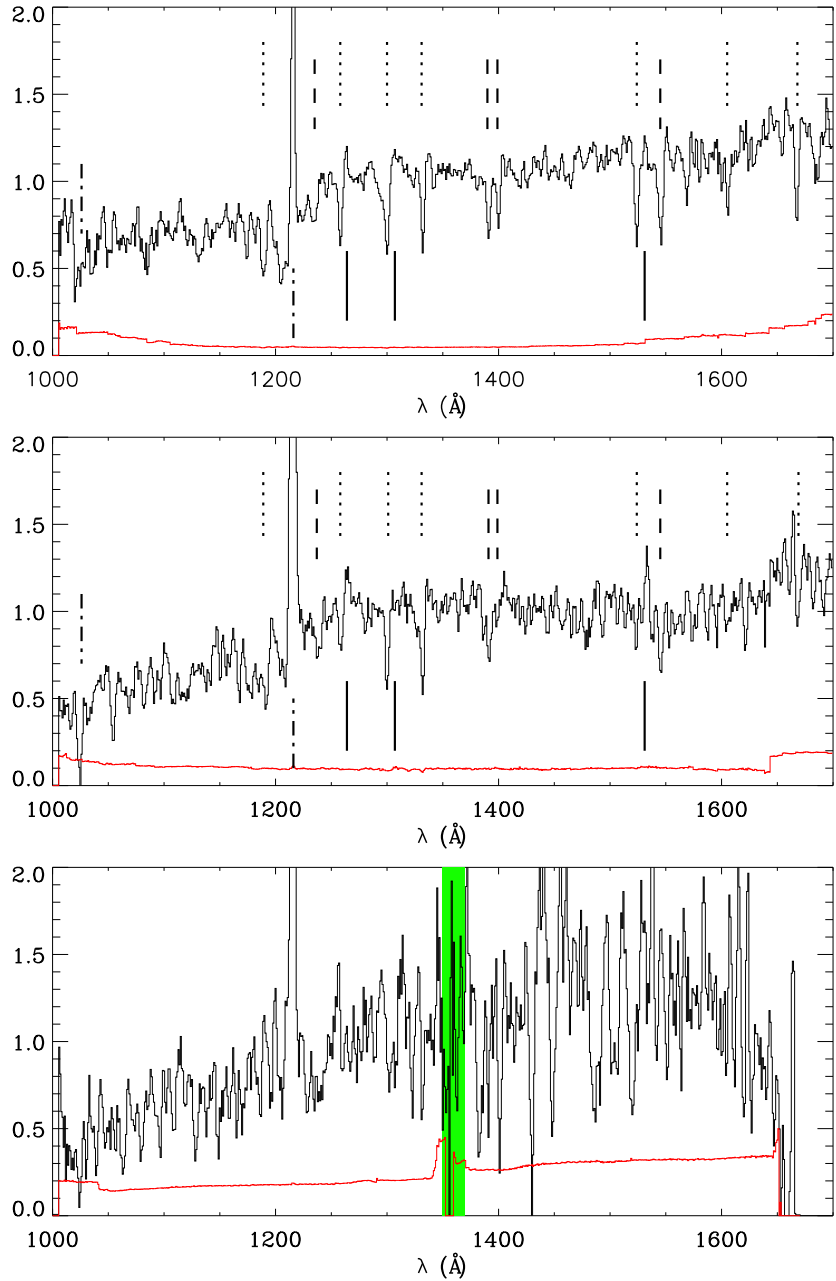}
    \caption{A series of continuum-normalized spectra for our three subsamples: top - UV-bright non-LAE, middle - UV-bright LAE, bottom - UV-faint LAE. The galaxies used to construct these spectra were shifted into the rest-frame using their Ly$\alpha$ emission line redshifts. The one-sigma composite error spectrum is plotted in red at the bottom of each composite. With increasing Ly$\alpha$ emission, interstellar absorption features become weaker, Si II$^{\ast}$ $\lambda \lambda$ 1265, 1309 emission becomes stronger, and the spectral slopes is bluer. The $\lambda 5577$ skyline has been masked in the UV-faint LAE composite (green) at $1360 {\mathrm \AA}$ as all UV-faint LAEs lie at $z=3.1$.}
\label{Fig:comp3_stack}
\end{figure}

In the UV-bright non-LAE composite (top panel of Fig.~\ref{Fig:comp3_stack}), we observe Ly$\alpha$ both in emission and blueshifted absorption with an average velocity offset of $\Delta v_{\mathrm {em-abs}}\sim2400$ km s$^{-1}$. The blueshifted absorption extends from the Ly$\alpha$ emission line to $\Delta v_{\mathrm {em-abs}} > 4500$ km s$^{-1}$ with an equivalent width of W$_o = -4.9\pm0.9 {\mathrm \AA}$. For the emission component, we measure a Ly$\alpha$ emission line equivalent width of W$_o = 8.4\pm0.9 {\mathrm \AA}$, which we find to be comparable to the second highest Ly$\alpha$ equivalent width quartile of S03 (W$_o = 11.00\pm0.71 {\mathrm \AA}$). Our results for the UV-bright non-LAE and UV-bright LAE composites are presented in Table \ref{Tab:comp3_prop}. The observed FWHM is right at our instrument resolution, so we set a 3$\sigma$ upper limit of FWHM(Ly$\alpha$) $<200$ km s$^{-1}$.

The UV-bright LAE composite (middle panel of Fig.~\ref{Fig:comp3_stack}) has a complex Ly$\alpha$ profile, as it includes galaxies with redshifted absorption and multiple-peaked emission. For the main Ly$\alpha$ emission line component, we measure deconvolved FWHM(Ly$\alpha$) = $410\pm100$ km s$^{-1}$ and an equivalent width of $W_o = 50.0\pm4.8 {\mathrm \AA}$. The faint blueshifted emission at very high velocity offset in this composite is the origin of the same feature in the UV-bright SFG composite. We measure a velocity offset of $\Delta v_{\mathrm {em-abs}}\sim4600$ km s$^{-1}$, which agrees with the maximum velocity offset of the Ly$\alpha$ absorption in the UV-bright non-LAE subsample. In the individual galaxy spectra contributing to the UV-bright LAE composite, we do not find any significant emission at this velocity offset, but we do note that the stacking procedure can reveal features too faint to be seen in individual spectra. The majority of Ly$\alpha$ photons will be absorbed at the relative velocity of the bulk of the outflowing neutral medium. However, Ly$\alpha$ photons farther in velocity space from the line center are less likely to be re-absorbed. \cite{Steidel2010} find outflows around $z=2.3$ SFGs to best be characterized by a powerlaw with larger velocities at larger radii. If the blueshifted emission at $\Delta v_{\mathrm {em-abs}}\sim4600$ km s$^{-1}$ is real, it reveals extended emission of Ly$\alpha$ photons from a diffuse component of H I. \cite{Steidel2011} recently found extended Ly$\alpha$ emission in a sample of 92 continuum-selected galaxies at z$\sim$2.65.

In our UV-faint LAE composite (bottom panel of Fig.~\ref{Fig:comp3_stack}), we see no strong evidence for blueshifted emission. We measure a Ly$\alpha$ equivalent width of $W_0 = 42.1\pm6.2 {\mathrm \AA}$ and a velocity FWHM of FWHM(Ly$\alpha$) = $380\pm110$ km s$^{-1}$, consistent with the UV-bright LAE composite.

        \subsection{Low-Ionization Absorption Lines}
\label{sec:LIS}
Blueshifted singly-ionized metal absorption lines probe the neutral component of outflowing interstellar material on the nearside of the galaxy. Among our composite spectra, there are a host of low-ionization interstellar absorption lines. Due to limited S/N, we restrict our analysis to the strongest low-ionization lines that we detect at high significance: Si II $\lambda 1260$, O I $\lambda 1302$ $+$ Si II $\lambda 1304$, C II $\lambda 1334$, Si II $\lambda 1526$. All of these features have been well-studied in high-redshift LBGs \citep{Pettini2000, Pettini2002, Shap2003, Cabanac2008, Quider2009, Dessauges-Zavadsky2010, Talia2011, Jones2011} and nearby starburst galaxies \citep{Heckman1998, Leitherer2011}. We also find Fe II $\lambda 1608$ and Al II $\lambda 1670$ at low significance, but they are located in a region of the spectrum where we have fewer galaxies contributing to the composite and therefore lower S/N. For these reasons, we only record their equivalent widths. Table \ref{Tab:comp3_prop} reports measurements of equivalent widths and average velocity offsets of the interstellar lines. As we observe a distribution of velocity offsets among our individual galaxies, the FWHMs of all transitions other than Ly$\alpha$ will be artificially broadened to some extent. Due to the intrinsically faint nature of our UV-faint LAE galaxies' continua in combination with continuum normalization, we do not observe any significant interstellar features in the UV-faint LAE composite. As a result, we restrict our analysis to the UV-bright LAE and UV-bright non-LAE composites for the rest of the analysis.

\begin{figure}[!t]
  \plotone{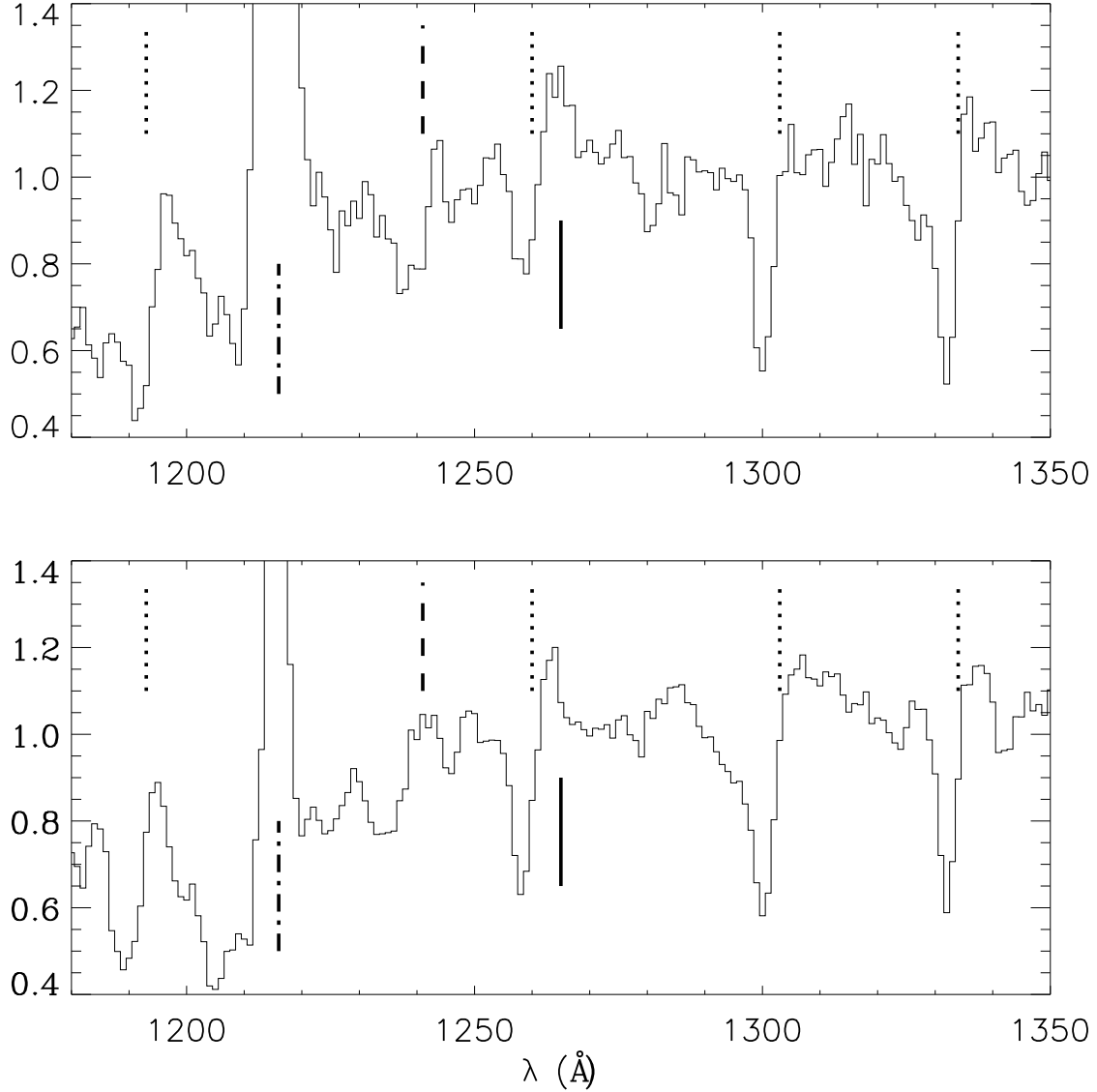}
    \caption{Zoomed-in regions of the UV-bright LAE (top panel) and UV-bright non-LAE (bottom panel) composite spectra shown in Figure 3. The adjusted scale allows for a more in depth look at the interstellar absorption line profiles as well as the nebular emission feature Si II$^{\ast}$ $\lambda \lambda$ 1265, 1309. Note the strength of the $\lambda$ 1265 emission as compared to the lack of $\lambda$ 1309 emission in both composites. The low-ionization transitions are all blueshifted with respect to Ly$\alpha$ and the N V $\lambda \lambda$ 1238, 1242 transition has a significantly higher velocity offset in the UV-bright non-LAE than LAE composite. The vertical lines indicate the Ly$\alpha$ rest-frame wavelengths with respect to Ly$\alpha$ emission for low-ionization absorption (dotted line), high-ionization absorption (dashed line), and Si II$^{\ast}$ emission (solid line). }
\label{Fig:LL_LB_IS1}
\end{figure}

\begin{figure}[!t]
  \plotone{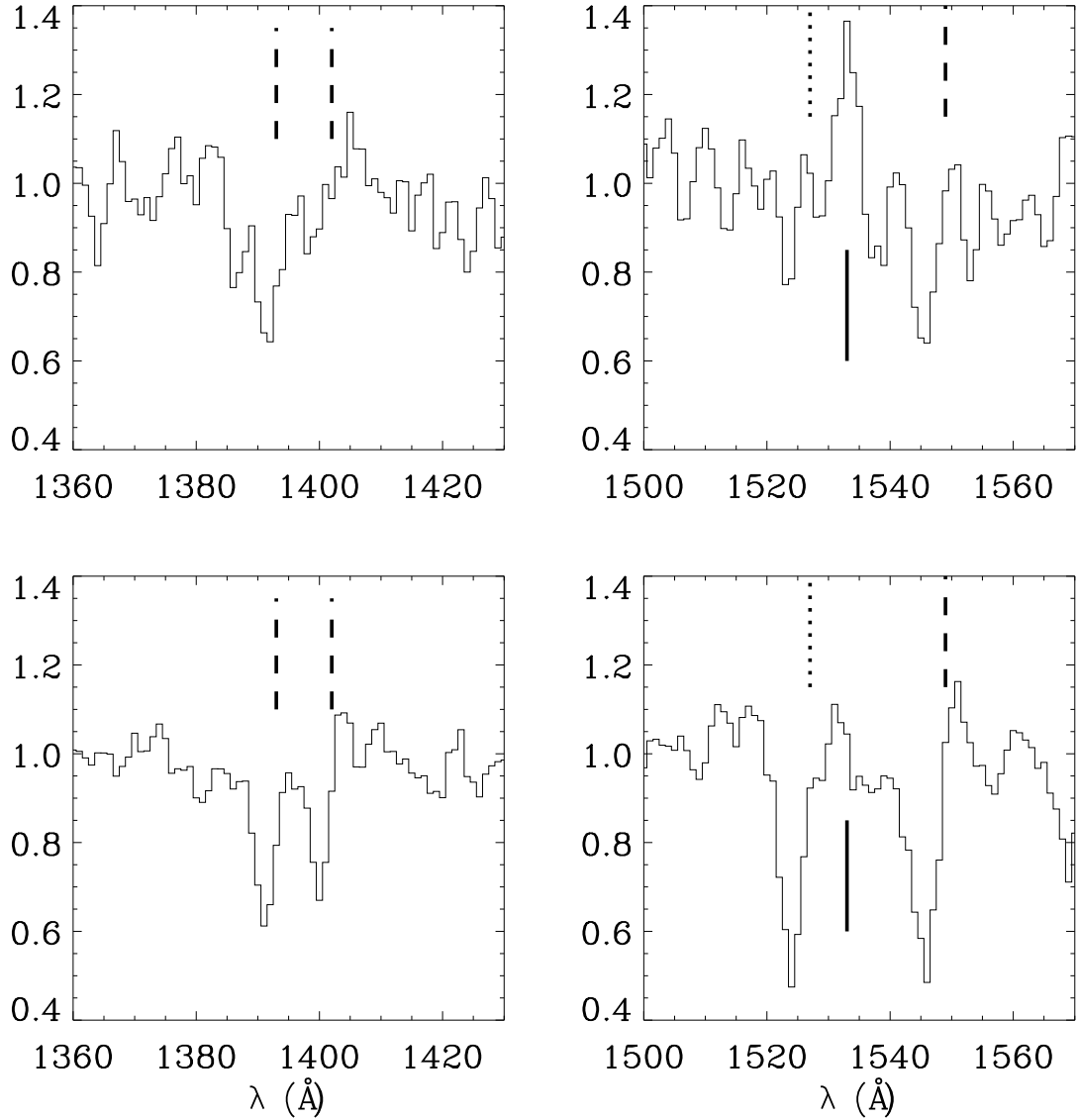}
    \caption{Analogous to Figure 4, except showing the Si IV $\lambda \lambda$ 1393, 1402, Si II $\lambda$1526 and C IV $\lambda \lambda$ 1548, 1550 transitions for the UV-bright LAE composite (top row) and UV-bright non-LAE composite (bottom row). This view shows the lack of Si IV $\lambda$1402 and Si II $\lambda$1526 absorption as well as the strength of the Si II$^{\ast}$ $\lambda$ 1533 emission in the UV-bright LAE composite. In the left panels, the Si IV redshifted emission is more apparent and the Si IV $\lambda \lambda$ 1393, 1402 absorption are seen at similar velocity offsets for each composite. In the right panels, the decrease in Si II$^{\ast}$ $\lambda$1526 emission and increase in Si II $\lambda$1526 and C IV $\lambda \lambda$ 1548, 1550 absorption with decreasing Ly$\alpha$ emission is clearer.}
\label{Fig:LL_LB_IS2}
\end{figure}

The strong low-ionization interstellar lines are useful for measuring the kinematics of the neutral gas component within the outflow. Figure \ref{Fig:LL_LB_IS1} and Figure \ref{Fig:LL_LB_IS2} show zoomed in regions around the strong absorption lines for the UV-bright LAE composite (top panel) and UV-bright non-LAE composite (bottom panel). The vertical lines indicate the rest-frame wavelength of each transition with respect to Ly$\alpha$ emission. Relative to Ly$\alpha$, we measure an average velocity offset of the low-ionization features of $\Delta v_{{\mathrm {em-abs}}} =640\pm10$ km s$^{-1}$ and $\Delta v_{{\mathrm {em-abs}}} =610\pm60$ km s$^{-1}$ for the UV-bright non-LAE and UV-bright LAE composites. Using the Si II $\lambda 1260$ and C II $\lambda 1334$ absorption lines, which are the least affected by blending, and assuming a spectral resolution of 3.0${\mathrm \AA}$, we measure a deconvolved FWHM = 430$\pm60$ km s$^{-1}$ for the UV-bright SFG composite. We subsequently measure a FWHM in the UV-bright LAE and UV-bright non-LAE composites of FWHM $=420\pm80$ km s$^{-1}$ and FWHM $=340\pm100$ km s$^{-1}$ respectively. The different velocity offsets in the UV-bright LAE and non-LAE composites is likely biasing the FWHM in the UV-bright SFG composite to larger values. Our results are consistent with S03, who stacked 811 LBGs at a spectral resolution of $2.6 {\mathrm \AA}$ and found an average FWHM(LIS) = $450\pm150$ km s$^{-1}$. 

We measure the degree of saturation of the Si II transitions by comparing the equivalent widths of Si II $\lambda 1260$ and Si II $\lambda 1526$. On the linear part of the curve of growth, the ratio of W$_{0}(1260)$/W$_{0}(1526) > 5$, while we measure W$_{0}(1260)$/W$_{0}(1526) = 1.1\pm0.2$ in the UV-bright SFG composite, consistent with unity and the material being optically thick. This ratio is W$_{0}(1260)$/W$_{0}(1526) = 2.4\pm1.5$ in the UV-bright LAE composite, as the $\lambda 1526$ transition is marginally detected, making the uncertainty quite large. S03, \cite{Erb2006}, and \cite{Steidel2010} find the Si II transitions to be consistent with saturation in each of their composites, and S03 report a W$_{0}(1260)$/W$_{0}(1526) = 0.95$ in their all-LBG composite. Since the strong low-ionization transitions are all saturated, they are not useful for determining chemical abundances. Metal abundances can be derived from weaker features lying on the linear part of the curve of growth in conjuction with H I column densities \citep[e.g.,][]{Pettini2000, Pettini2002}. We do not observe any of these features with high enough S/N and cannot measure H I column densities, so we focus our analysis on the strong transitions. We do note that \cite{Erb2006z} used the metallic absorption lines near $\lambda$1370 and $\lambda$1425 to confirm their rest-frame optical metallicity measurements. We find these regions in our UV-bright SFG composite to be consistent with the low-mass composite of \cite{Erb2006z} which had a metallicity of $Z<0.33Z_{\odot}$. \cite{Finkelstein2009} measured limits on the metallicities of two LAEs using rest-frame optical emission lines to be $Z<0.17Z_{\odot}$ and $Z<0.28Z_{\odot}$.

        \subsection{High-Ionization Absorption Lines}
\label{sec:HIS}
In addition to low-ionization absorption, we also detect at high significance several high-ionization interstellar absorption lines, namely N V $\lambda \lambda 1238, 1242$, Si IV $\lambda \lambda 1393, 1402$, and C IV $\lambda \lambda 1548, 1550$. These transitions trace gas at T $\ge 10^4$ K that has been ionized by feedback from supernovae, stellar winds, and collisional processes associated with the outflow. They are therefore useful for probing the ionized component and ionization state of the outflow. Among all of our composites, we find that the high-ionization line profiles are dominated by the interstellar component with a smaller contribution from the stellar component. We analyze the most prominent high-ionization features first (Si IV and C IV), then interpret the N V line profile, which we detect at lower significance.

In the UV-bright non-LAE composite, for the $\lambda$$\lambda$ $1393,1402$ Si IV transitions, we measure comparable velocity offsets of $\Delta v_{{\mathrm {em-abs}}} \sim 560$ km s$^{-1}$ and $\Delta v_{{\mathrm {em-abs}}} \sim 590$ km s$^{-1}$. In contrast, for the UV-bright LAE composite, we measure velocity offsets of $\Delta v_{{\mathrm {em-abs}}} = 440$ km s$^{-1}$ and $\Delta v_{{\mathrm {em-abs}}} = 870$ km s$^{-1}$ respectively. The UV-bright LAE composite shows strong absorption at $\lambda 1393$ while having marginally detected absorption at $\lambda 1402$ likely due to redshifted emission from a P Cygni component filling it in, and thereby biasing our velocity offset to a higher value. We also find the signature of a P Cygni profile in the UV-bright non-LAE composite, although at lower significance and it does not appear to have a significant effect on the velocity offset of the $\lambda$1402 absorption. For a doublet ratio of 2:1, the Si IV $\lambda$$\lambda$ $1393,1402$ transition is on the linear part of the curve of growth and therefore optically thin. In the UV-bright SFG composite, this ratio is 1.9:1, however, this ratio is 3.5:1 in the UV-bright LAE composite and 1.4:1 in the UV-bright non-LAE composite. The higher doublet ratio in the UV-bright LAE composite is again due to P Cygni emission filling in $\lambda$1402 absorption. In the UV-bright non-LAE composite, continuum uncertainty due to broad stellar absorption makes measuring this ratio difficult and may bias this ratio towards 1:1. Thus, we conclude that the Si IV transition appears consistent with being optically thin in all composites.

For the UV-bright SFG composite, we find the relative velocity offset of the Si IV $\lambda$$\lambda$ 1393, 1402 transitions ($\Delta v_{{\mathrm {em-abs}}} \sim 590$ km s$^{-1}$) to be in good agreement with the average velocity shift of the low-ionization absorption lines ($\Delta v_{{\mathrm {em-abs}}} = 550$ km s$^{-1}$). Moreover, the average deconvolved FWHM of the Si IV doublet is FWHM$_{Si IV}=420\pm110$ km s$^{-1}$, also consistent with that of the low-ionization absorption lines FWHM$_{LIS}=360\pm60$ km s$^{-1}$.

In the UV-bright SFG composite, for C IV blended at $\lambda 1549.5$, assuming a doublet ratio of 1:1, we measure $\Delta v_{{\mathrm {em-abs}}} = 800$ km s$^{-1}$, significantly larger than the low-ionization lines. However, the velocity width of FWHM$_{C IV}=450\pm60$ km s$^{-1}$ is consistent with the low-ionization lines. There appears to be some broadening due to stellar winds along with redshifted emission characteristic of a P Cygni profile, which may be biasing the velocity offset to larger values.  For both the UV-bright LAE and UV-bright non-LAE composites, we find comparable velocity offsets to the UV-bright SFG composite, which are in all cases larger than that of the Si IV $\lambda 1393$ transition and the low-ionization absorption lines.

In the UV-bright LAE composite, we see weak C IV absorption, yet in the UV-bright non-LAE composite, we find significantly stronger C IV absorption corresponding to an increase in absorption strength of $\sim$100\%. S03 also report seeing more C IV absorption in their W$_{\mathrm {Ly\alpha}}=11{\mathrm \AA}$ composite than their W$_{\mathrm {Ly\alpha}}=52{\mathrm \AA}$ composite, although they only measure an increase in strength of $\sim$30\%. The relatively weak C IV absorption seen in the UV-bright LAE composite is particularly interesting given that the strength of the Si IV $\lambda 1393$ absorption is comparable in both the UV-bright LAE and UV-bright non-LAE composites, and the velocity FWHMs of the high-ionization transitions are comparable across all composites. We further note from the relative strength of the interstellar C II $\lambda 1334$ absorption to C IV $\lambda$1549 absorption that the outflowing component of carbon appears to have a higher covering fraction in a lower ionization state, as it is unlikely that the velocity FWHM of the neutral gas is higher than that of the ionized gas.

Significant N V absorption is visible in the UV-bright SFG composite with a large velocity FWHM (FWHM$\sim1200$ km s$^{-1}$) and velocity offset ($\Delta v_{\mathrm {em-abs}}=1150$ km s$^{-1}$). The N V line profile shape is consistent with that of a P Cygni profile, indicating that there is both a stellar and an interstellar component. The velocity offset of the N V absorption is significantly larger in the UV-bright non-LAE composite ($\Delta v_{{\mathrm {em-abs}}}=1300$ km s$^{-1}$) than the UV-bright LAE composite ($\Delta v_{{\mathrm {em-abs}}}=600$ km s$^{-1}$). In spite of this difference in velocity offset, the velocity width is consistent among all three composites (FWHM$\sim1200$ km s$^{-1}$). We interpret the larger velocity offset in the UV-bright non-LAE composite as evidence for a significant amount of interstellar N V absorption in the outflowing component. In this composite, both the N V velocity offset and velocity FWHM are significantly larger than that of the Si IV and C IV absorption.

        \subsection{Stellar Features}
\label{sec:stellar_features}
The radiation pressure from hot massive stars can generate winds of 2000-3000 km s$^{-1}$ \citep{Groenewegen1989}. These winds are seen spectroscopically as broad absorption in low density environments or P Cygni profiles, blueshifted absorption followed by redshifted emission, in higher density environments \citep{Leitherer1995}. In the far UV, the most prominent stellar features are N V $\lambda \lambda 1238$, 1242, Si IV $\lambda \lambda$ 1393, 1402, C IV $\lambda \lambda$ 1548, 1550, and He II $\lambda$1640. There is no evidence for He II $\lambda 1640$ emission in any of the composites, however, the S/N in that region is much lower due to limited spectral coverage and an uncertain continuum level. For the Si IV and C IV transitions, a combination of strong interstellar and photospheric absorption contribute to the line profiles, which can be difficult to differentiate. In the UV-bright LAE composite, we find evidence for O III] $\lambda 1664$ nebular emission, which is not detected in the UV-bright non-LAE composite. Due to low S/N and an uncertain continuum level, we do not attempt any further analysis. This result is qualitatively consistent with S03 who reported significantly stronger O III] $\lambda 1664$ emission in their highest Ly$\alpha$ equivalent width quartile relative to their lower three Ly$\alpha$ equivalent width quartiles. A number of weak stellar features are also visible in the UV-bright SFG composite, the strongest being C III $\lambda 1176$ absorption at a velocity offset of $\Delta v_{\mathrm {em-abs}}=380$ km s$^{-1}$, which further confirms that Ly$\alpha$ is typically redshifted from the systemic redshift of SFGs at these redshifts by $\sim$400 km s$^{-1}$ in excellent agreement with \cite{Steidel2010}.

The stellar wind component of the Si IV transition is only significant for blue giant and supergiant stars. In the UV-bright SFG composite, we observe extended absorption that ranges from zero velocity offset to $\Delta v_{{\mathrm {em-abs}}}=1600$ km s$^{-1}$ as well as redshifted emission characteristic of a P Cygni profile, consistent with high velocity stellar winds from massive stars. We observe a similar profile in the UV-bright LAE composite except with redshifted emission filling in the Si IV $\lambda$1402 absorption. Although the UV-bright non-LAE composite shows weak redshifted emission, there does not appear to be significant broad absorption. Given the strength of Si IV $\lambda$ 1393, the most significant difference between these two composites is the much lower amount of Si IV $\lambda 1402$ interstellar absorption in the UV-bright LAE composite. The shape and strength of these line profiles indicates the presence of a population of young blue giant and supergiant stars.

The stellar wind contribution to the C IV profile is most dependent on the presence of main-sequence, giant and supergiant O stars. We observe strong C IV $\lambda \lambda$ 1548, 1550 absorption in the UV-bright non-LAE composite, though it is difficult to disentangle the stellar wind component from the interstellar absorption component. We find evidence for weak extended absorption in the UV-bright SFG and UV-bright non-LAE composites, but not in the UV-bright LAE composite. We also find significant redshifted emission in the UV-bright non-LAE composite. A P Cygni profile indicates the presence of M$\ge$30M$_{\odot}$ stars \citep{Leitherer1995, Pettini2000}. This emission is also seen in the UV-bright SFG composite.

In both the UV-bright LAE and UV-bright SFG composites, there is significant blueshifted absorption and redshifted emission in the N V line profile, characteristic of a P Cygni profile. We interpret the low velocity offset ($\Delta v_{{\mathrm {em-abs}}}=600$ km s$^{-1}$) and large velocity FWHM (FWHM$\sim1200$ km s$^{-1}$) of N V absorption in the UV-bright LAE composite as evidence for stellar emission and absorption dominating the N V profile \citep[see Fig.~1 in][]{Groenewegen1989}. Given the shape of the N V line profile in the UV-bright SFG composite, we interpret this as evidence for a significant amount of high-ionization absorption in the outflow as well as emission and absorption due to stellar features at the systemic redshift.

        \subsection{Si II$^{\ast}$ Emission Lines}
\label{sec:SiII}
An advantage of composite spectra is the ability to reveal weak features that are typically undetected in individual spectra. Among these features, the most prominent are the Si II$^{\ast}$ fine structure emission lines at $\lambda 1265$, $\lambda 1309$, and $\lambda 1533$. These transitions have also been observed in z$\sim$3 LBGs by S03, z$\sim$4 LBGs by \cite{Jones2011}, and in a local LBG by \cite{France2010}.

The close proximity of these lines to strong absorption features (Si II $\lambda 1260$, O I $+$ Si II $\lambda 1303$, and Si II $\lambda 1526$) may attenuate the blue edges, biasing the centroids to the red. Our results for Si II$^{\ast}$ emission in the UV-bright SFG composite are presented in Table \ref{Tab:SiII}. In the UV-bright non-LAE composite, we observe significant Si II$^{\ast}$ $\lambda 1265$ emission yet we do not see the $\lambda 1309$ and $\lambda 1533$ transitions. For the $\lambda$1265 emission, we measure an equivalent width of W$_{0} = 0.5\pm0.2 {\mathrm \AA}$ at a velocity offset of $\Delta v_{{\mathrm {em-abs}}}=300$ km s$^{-1}$. We find slightly stronger Si II$^{\ast}$ $\lambda$1265 in the UV-bright LAE composite with an equivalent width of W$_{0}=0.7\pm0.2 {\mathrm \AA}$ at a velocity offset of $\Delta v_{{\mathrm {em-abs}}}=120$ km s$^{-1}$. In this composite, the Si II$^{\ast}$ $\lambda$1265 emission is double-peaked, which is likely due to only individual galaxies having Si II$^{\ast}$ emission at different velocity offsets. We observe $\lambda$1533 emission at a much higher signficance in the UV-bright LAE composite than the UV-bright non-LAE composite with an equivalent width of W$_{0}=1.2\pm0.4 {\mathrm \AA}$ compared to W$_{0}<0.3 {\mathrm \AA}$ at velocity offsets of $\Delta v_{{\mathrm {em-abs}}}=30 {\mathrm km s^{-1}}$ and $\Delta v_{{\mathrm {em-abs}}}=380 {\mathrm km s^{-1}}$ respectively. We do not find any evidence for $\lambda$1309 emission in any of the composite spectra. In contrast, both S03 and \cite{Jones2011} observe Si II$^{\ast}$ $\lambda$1309 emission in their composite spectra.

Si II$^{\ast}$ emission likely stems from photoionized regions around massive stars in the systemic component of the galaxy where electron densities are $N_e = 10^2 - 10^3$ cm$^{-3}$ with temperatures of $T=10^{4}$K. In this environment, the Si II and Si III relative abundances are determined from radiative recombinations for which the recombination rates are comparable to the collisional excitation rates \citep{Shull1982}. S03 find, through modeling the LBG nebular emission lines with the CLOUDY96 software package, that all models that simultaneously match their O, C, and H$\beta$ line ratios overpredict the Si II emission-line strengths by more than an order of magnitude. This result leads them to conclude that the Si II emission lines cannot originate from H II regions. As we do not observe these other transitions, we are unable to comment on the plausibility of Si II$^{\ast}$ fine structure emission stemming from H II regions.

A different explanation for Si II$^{\ast}$ emission is that these lines originate instead from the outflowing component. In this scenario, each UV photon absorbed at a resonance transition ($\lambda = 1260, 1304, 1526$) or a fine-structure transition must be re-emitted through either a resonance or a fine-structure transition. In the absence of dust, the sum of the resonance and fine-structure emission should equal the amount of absorption. Our UV-bright non-LAE composite has significantly stronger resonance absorption than fine-structure emission while our UV-bright LAE composite shows a much larger ratio of fine-structure emission to resonant absorption. In contrast, S03 observe the fine-structure emission lines to be an order of magnitude weaker than the resonance lines. The discrepancy between the emission and absorption components could be due to the presence of dust extinguishing Si II photons during resonant scattering. Furthermore, we observe the Si II$^{\ast}$ fine-structure lines to have narrow velocity widths, whereas we would expect the emission to come from the full range of outflow velocities and to therefore be as broad as the Ly$\alpha$ emission line. For these reasons, we find it unlikely that the Si II fine-structure lines originate from the outflowing component. \cite{Erb2010} also discuss possible mechanisms for the origin of fine-structure emission lines, but find no satisfactory explanation.

\section{Discussion}
\label{sec:discussion}

Combining the spectra of both continuum- and narrowband-selected galaxies has allowed us to study the average properties of $2<z<3.5$ SFGs. Among these properties, we are able to examine trends based on Ly$\alpha$ equivalent width, rest-frame UV color corrected for IGM extinction, redshift, and R band luminosity for each individual galaxy. Due to the complexities associated with selection effects, we do not attempt to draw any inferences from different R band luminosities. We have divided our sample based upon Ly$\alpha$ equivalent width for galaxies with R$<$25.5 and will first examine trends associated with different amounts of Ly$\alpha$ emission. Due to the faint nature of our objects' continua, we parameterize their observed colors in terms of their photometrically measured rest-UV spectral slope after correcting for IGM extinction and Ly$\alpha$ equivalent width, then divide our sample in half based on this parameter. Additionally, our spectroscopic sample divides evenly into two redshift bins ($2<z<2.7$, $2.7<z<3.5$) allowing us to look for any trends across this redshift interval. This allows us to study trends in spectroscopic properties in relation to spectral slope and redshift.

        \subsection{Selection Effects}
\label{sec:sel_eff}
As our subsamples of galaxies are determined on the basis of different selection criteria, it is important to identify any photometric or spectroscopic biases that may affect our analysis of the underlying galaxy populations. For continuum-selected galaxies, one of the most influential effects is the dependence of color selection criteria on luminosity and redshift due to Ly$\alpha$ forest absorption. Another major bias stems from the prevalence of spectroscopic redshifts determined solely from Ly$\alpha$ emission in conjunction to a minimum Ly$\alpha$ flux necessary for Ly$\alpha$ emission to be seen. For galaxies without distinct absorption lines, this causes spectroscopic identification to be Ly$\alpha$ flux dependent.

The selection process for our continuum-selected objects creates two major biases that affect our resultant galaxy population. First of all, our continuum-selected LBGs were chosen to have V-R $<$ 1.2 thereby targeting objects within a redshift range of $2.6<z<3.5$. However, the V band is extinguished by the Ly$\alpha$ forest making it redshift dependent. At $z=2.5$, the amount of IGM extinction in the V band is $\Delta$V$=0.0$ mag, while at $z=3.5$, $\Delta$V$=0.5$ mag. For this reason, the range of intrinsic colors of the continuum-selected population is a function of redshift. Secondly, the colors of continuum-selected galaxies is also a function of R band magnitude. The combination of requiring our targets to have very faint U band magnitudes ($(U-V)_{AB}>1.2$) and U band images having a finite depth will indirectly restrict the range of V-R color to bluer colors at fainter luminosities.

\begin{figure}[!t]
  \plotone{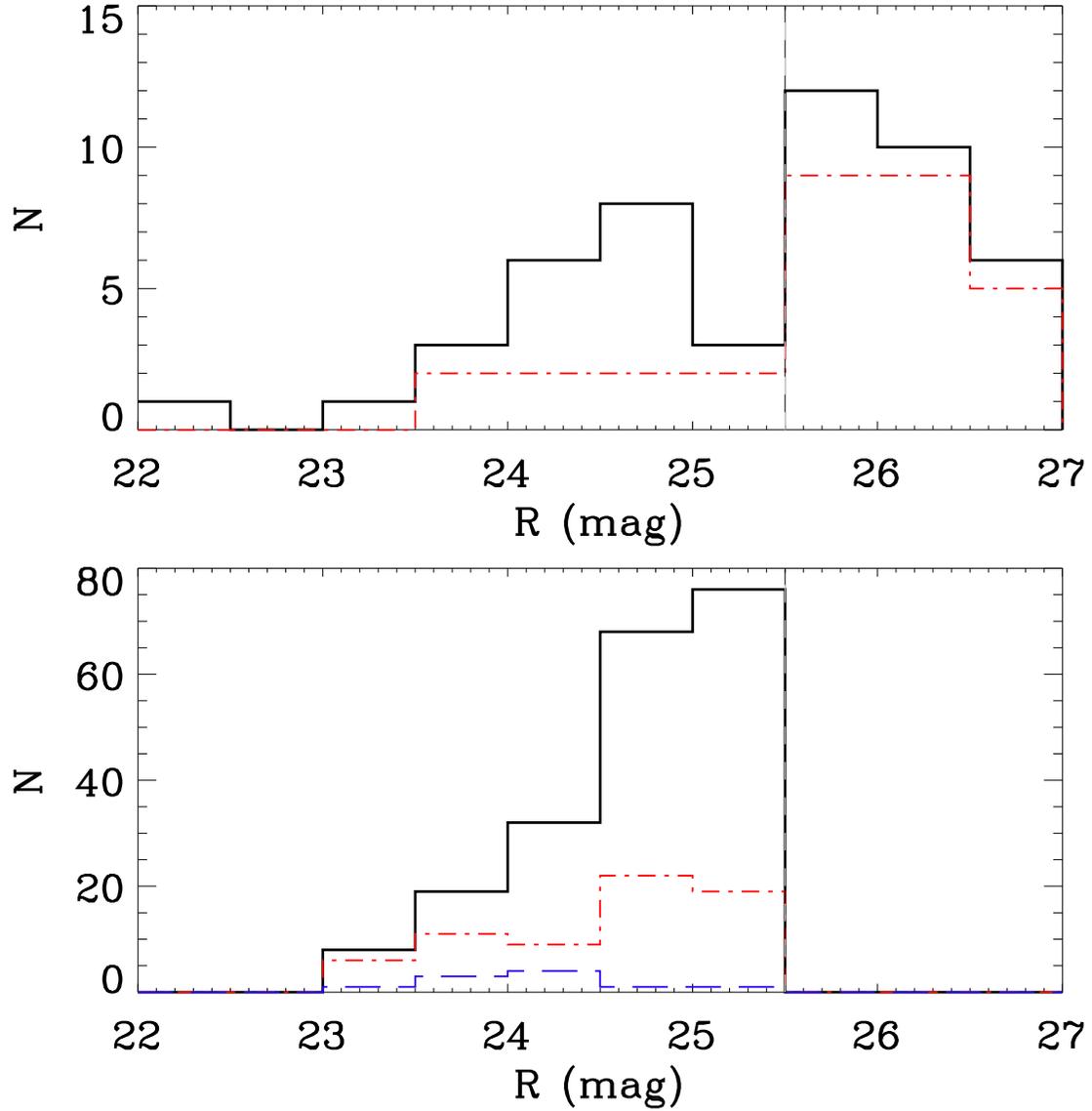}
    \caption{Spectroscopic incompleteness as a function of R band magnitude and spectral type is plotted for the entire catalog. Top panel - the black solid histogram indicates the total number of narrowband-selected objects targeted spectroscopically and the red dot dashed line shows the number that were spectroscopically confirmed. Bottom panel -  the black solid histogram indicates the total number of broadband-selected objects targeted spectroscopically, the red dot dashed line shows the number that were spectroscopically confirmed, and the blue dashed line shows the number that were spectroscopically confirmed galaxies without Ly$\alpha$ in emission. Our spectroscopic confirmation rate is 33\% for broadband-selected objects and 60\% for narrowband-selected objects. The vertical dashed line represents the canonical $R\le 25.5$ cut for broadband-selected galaxies.}
\label{Fig:spec_rate}
\end{figure}

Spectroscopic success rate is not only dependent on apparent magnitude, but also on the fact that it is easier to measure a redshift from an emission line rather than an absorption line. Our spectroscopic incompleteness as a function of R band magnitude is plotted in Fig \ref{Fig:spec_rate} for narrowband-selected objects (top panel) and broadband-selected objects (bottom panel). In each panel, the black line indicates the total number of objects observed spectroscopically and the red dot dashed line shows the number of spectroscopically confirmed galaxies. The blue dashed line shows the number of spectroscopically confirmed galaxies without Ly$\alpha$ emission. Our confirmation rates are 63\% for narrowband-selected objects and 33\% for broadband-selected objects. Only 5\% of broadband-selected objects have redshifts determined solely from absorption lines. Due to the low S/N of our individual spectra, we found it difficult to robustly measure redshifts for galaxies without Ly$\alpha$ emission. Galaxies without Ly$\alpha$ emission line redshifts likely have Ly$\alpha$ equivalent widths below a certain threshold as a function of luminosity. We assume this to be consistent with no Ly$\alpha$ emission. \cite{Steidel2000} find that roughly half of their z$\sim$3 spectroscopically-confirmed LBGs show Ly$\alpha$ in absorption. Assuming a similar distribution among our broadband-selected objects, this would imply a spectroscopic confirmation rate of 56\%.

        \subsection{Individual Galaxy Ly$\alpha$ Profiles}
\label{sec:ind_lya_prop}

Recently, \cite{Laursen2011} used cosmological hydrosimulations to examine the effect of the IGM on the Ly$\alpha$ line profile escaping high-redshift, star-forming galaxies. Their findings qualitatively agree on the shape and strength of the redshifted Ly$\alpha$ emission line, however, they predict more blueshifted emission than has typically been observed for high-redshift SFGs. At a redshift of $z=3.5$, they find the effect of the IGM is insufficient to explain the lack of a blueshifted Ly$\alpha$ emission line. For our sample of 12 multiple-peaked Ly$\alpha$ emission line galaxies, we find a mean velocity offset of $<\Delta v_{{\mathrm red-blue}}>=660\pm229$ km s$^{-1}$. This measurement is in good agreement with \cite{Kulas2011} who found a mean velocity offset of $<\Delta v_{{\mathrm red-blue}}>=741\pm39$ km s$^{-1}$ (errors represent the standard deviation of the mean). However, we do find a larger range of velocity offsets than \cite{Kulas2011} from $\Delta v_{{\mathrm red-blue}}=330$ km s$^{-1}$ to $\Delta v_{{\mathrm red-blue}}=1080$ km s$^{-1}$. We note that this result is consistent with \cite{Steidel2010} who found a broad range of Ly$\alpha$ emission line velocity offsets ($\Delta v=+485\pm175$ km s$^{-1}$) with respect to the systemic redshift. This distribution is also comparable to the observed velocity offset between Ly$\alpha$ emission and interstellar absorption seen in individual spectra (Fig.~\ref{Fig:ind_gal_prop}).

Among the spectroscopic sample, there is a broad distribution of Ly$\alpha$ line strengths and profile types, which fall into four categories: emission only, absorption only, both emission and absorption, and neither emission nor absorption. Out of the continuum-selected galaxies in our composites, we find 45\% have Ly$\alpha$ emission lines strong enough to be classified as LAEs (W$_{\mathrm {Ly\alpha}} > 20{\mathrm \AA}$) while only 10\% show Ly$\alpha$ only in absorption. This fraction of UV-bright LAEs is roughly a factor of two larger than the $20-25\%$ that S03 report for their z$\sim$3 LBGs. However, due to the poor S/N of our data, it was difficult to determine redshifts for LBGs without Ly$\alpha$ in emission and we remove any galaxies lacking precise redshifts. Removing all galaxies without Ly$\alpha$ emission, our fraction of LBGs that would be classified as LAEs rises to 50\% and the fraction of LBG-LAEs from \cite{Steidel2000} increases to $\sim$40\%. Excluding LBGs with Ly$\alpha$ in absorption, we report a median Ly$\alpha$ equivalent width of W$_{Lya}=20.0 {\mathrm \AA}$. Similarly, two of our R$<$25.5 narrowband-selected objects did not yield spectroscopic Ly$\alpha$ equivalent widths high enough to be classified as LAEs and were included in the UV-bright non-LAE stack. We find that 24\% of our continuum-selected galaxies have both Ly$\alpha$ in emission and absorption. For these objects, we use their emission line equivalent widths in our analysis.

        \subsection{Composite Ly$\alpha$ Dependences}
\label{sec:Lya_dep}

\begin{figure}[!t]
  \plotone{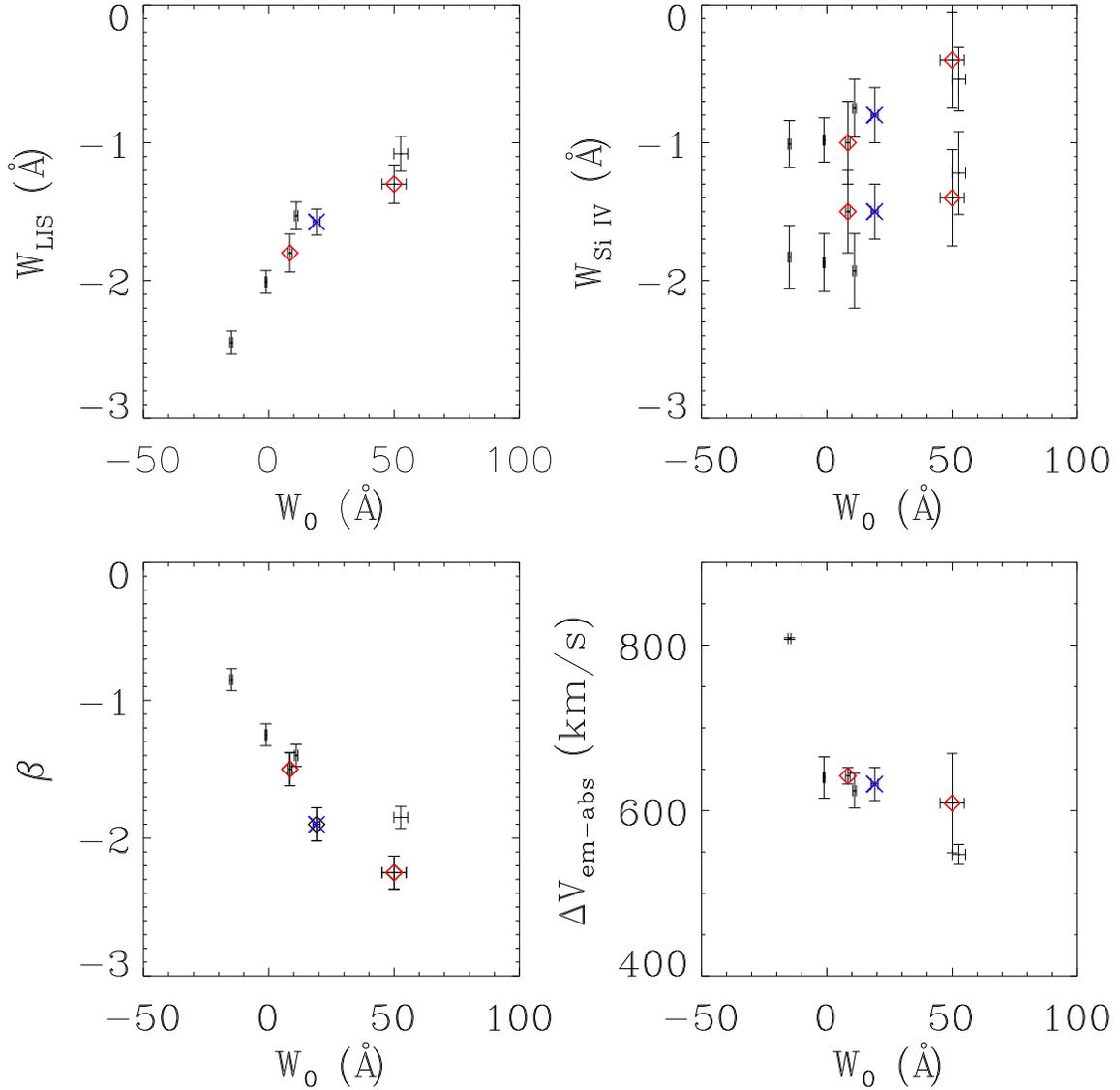}
    \caption{Dependences of UV spectroscopic properties on Ly$\alpha$ equivalent width for the UV-bright non-LAE and UV-bright LAE subsamples as red diamonds, the UV-bright SFG composite in blue and the four Ly$\alpha$ equivalent width quartiles from S03 in black. Top left - average low-ionization equivalent widths (Si II $\lambda$ 1260, O I + Si II $\lambda \lambda$ 1302,1304, C II $\lambda$ 1334, Si II $\lambda$ 1526). Top right - Si IV equivalent widths. Bottom left - spectral slope, $\beta$, as measured from the composite spectra. The error bars for our $\beta$ values reflect the error in spectral slope index and do not take into account the uncertainty in the artificial flux calibration of the VIMOS spectra. Bottom right - velocity offset $\Delta v_{{\mathrm {em-abs}}}$. The errors in velocity offset represent the range of velocity offsets for different low-ionization lines. The trends in spectroscopic properties with increasing Ly$\alpha$ emission originally observed by S03 are confirmed for our smaller sample of SFGs spanning a larger range in redshift.}
\label{Fig:Lya_dep}
\end{figure}

Through dividing our bright galaxies (R$<$25.5) into a UV-bright non-LAE and a UV-bright LAE composite, we can look for trends based on Ly$\alpha$ equivalent width. The clearest trend across these two composites is stronger interstellar absorption with decreasing Ly$\alpha$ emission. The top panel of Figure \ref{Fig:Lya_dep} shows the highest Ly$\alpha$ equivalent width quartile of S03 (top, red; W$_{\mathrm {Ly\alpha}}=52.63{\mathrm \AA}$) and the UV-bright LAE composite (bottom, black; W$_{\mathrm {Ly\alpha}}=50.0{\mathrm \AA}$) while the bottom one shows the second highest Ly$\alpha$ equivalent width quartile of S03 (top, red; W$_{\mathrm {Ly\alpha}}=11.00{\mathrm \AA}$) and the UV-bright non-LAE composite (bottom, black; W$_{\mathrm {Ly\alpha}}=8.4{\mathrm \AA}$). Across these two pairs of composite spectra, there are many interesting differences including Si II$^{\ast}$ emission, low- and high-ionization absorption strength, UV spectral slope, and velocity offset. In both the UV-bright LAE and UV-bright non-LAE composites, we find stronger Si II$^{\ast}$ emission at $\lambda$1265 and weaker Si II$^{\ast}$ emission at $\lambda$1309 than the W$_{\mathrm {Ly\alpha}}=52.63{\mathrm \AA}$ and W$_{\mathrm {Ly\alpha}}=11.00{\mathrm \AA}$ composites of S03. Furthermore, for both our composite spectra and that of S03, Si II$^{\ast}$ emission becomes stronger with increasing Ly$\alpha$ emission.

\begin{figure}[!t]
\epsscale{0.8}
  \plotone{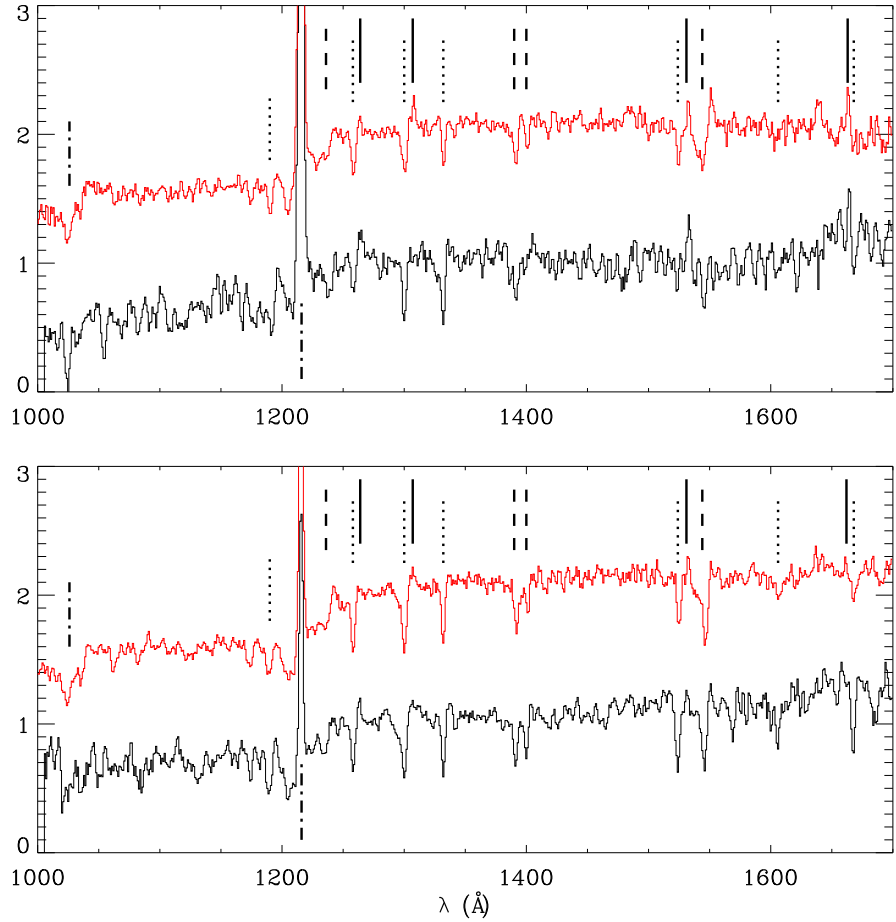}
    \caption{Top - the highest Ly$\alpha$ equivalent width quartile composite (W$_{\mathrm {Ly\alpha}}=52.63 {\mathrm \AA}$) from S03 in red and the UV-bright LAE composite (W$_{\mathrm {Ly\alpha}}=50.0 {\mathrm \AA}$) in black. Bottom -  the second highest Ly$\alpha$ equivalent width quartile composite (W$_{\mathrm {Ly\alpha}}=11.00 {\mathrm \AA}$) from S03 in red and the UV-bright non-LAE composite (W$_{\mathrm {Ly\alpha}}=8.4 {\mathrm \AA}$) in black. Our composite spectra show similar trends in both interstellar absorption and Si II$^{\ast}$ emission as those of S03. In the top panel, S03 find stronger Si II$^{\ast}$ $\lambda$1309 emission, weaker Si IV $\lambda$1402 nebular emission, and a more prominent C IV $\lambda \lambda$ 1548, 1550 profile. Note that the average low-ionization equivalent widths for the four S03 quartiles are different from figure 9 of S03 as we do not include the Fe II $\lambda$1608 and the Al II $\lambda$ 1670 transitions in our measurement.}
\label{Fig:B_SH_ind_stack}
\end{figure}

Due to poor S/N in the $\lambda > 1600 {\mathrm \AA}$ continuum, we only use the bluest low-ionization absorption lines ($\lambda 1260, \lambda 1303, \lambda 1334, \lambda 1527$) in determining the average low-ionization line equivalent widths. Figure \ref{Fig:B_SH_ind_stack} shows the dependences of various spectroscopic features on Ly$\alpha$ equivalent width. As can be seen in the top left panel of Figure \ref{Fig:B_SH_ind_stack}, the average low-ionization equivalent width decreases from W$_{LIS}=-1.8 {\mathrm \AA}$ to W$_{LIS}=-1.25 {\mathrm \AA}$ for a change in Ly$\alpha$ equivalent width of W$_0=8.4 {\mathrm \AA}$ to W$_0=50.0 {\mathrm \AA}$. We find a similar trend in low-ionization absorption strength with respect to Ly$\alpha$ equivalent width as S03. Note that the average low-ionization equivalent widths for the four S03 quartiles are different from figure 9 of S03 as we do not include the Fe II $\lambda$1608 and the Al II $\lambda$ 1670 transitions in our measurement. For the same set of transitions, we find systematically stronger low-ionization absorption at a given Ly$\alpha$ equivalent width. In the UV-bright LAE composite, the Si II $\lambda 1526$ transition is only marginally detected, which differs from the highest Ly$\alpha$ equivalent width quartile in S03. The degree of saturation of Si II can be measured from the ratio of Si II $\lambda 1260$ to Si II $\lambda 1526$, which is optically thin for W$_{0}(1260)$/W$_{0}(1526)>5$. The ratio of Si II absorption line equivalent widths is W$_{0}(1260)$/W$_{0}(1526)=0.8\pm0.2$, consistent with unity in the UV-bright non-LAE composite while it is W$_{0}(1260)$/W$_{0}(1526)=2.4\pm1.5$ in the UV-bright LAE composite. While this may indicate that Si II is not optically thick in the UV-bright LAE composite, the faint nature of both transitions makes this ratio uncertain. With increasing Ly$\alpha$ equivalent width, we observe an average decrease of $>50\%$ in the strength of the Si II $\lambda$$\lambda$ 1260, 1526 and O I + Si II $\lambda$1303 absorption. However, we note that the strength of C II $\lambda$1334 instead increases by $20\%$ with larger Ly$\alpha$ equivalent width.

Robustly determining Si IV equivalent widths is challenging given the uncertainty in the continuum level, which is affected by broad stellar absorption as well as redshifted emission. In the top right panel of Figure \ref{Fig:B_SH_ind_stack}, Si IV $\lambda$$\lambda$ 1393, 1402 equivalent widths are plotted for this work (red diamonds) and S03 (black points). The UV-bright LAE Si IV $\lambda$1402 equivalent width measurement is likely being filled in by redshifted emission, and the $\lambda$1393 transition is the same strength as in the UV-bright non-LAE composite. For this reason, we find no significant change in Si IV equivalent widths and conclude that the transition is consistent with being optically thin, which for Si IV corresponds to a doublet ratio of W$_{1393}$/W$_{1402}=2$. Furthermore, we do observe a significantly smaller C IV absorption strength in the UV-bright LAE composite $W_{CIV}=-1.1\pm0.35$ than the UV-bright non-LAE composite $W_{CIV}=-2.2\pm0.4$, corresponding to a decrease of $\sim100\%$. S03 reported finding $50\%$ weaker high-ionization equivalent widths (Si IV and C IV) in their highest Ly$\alpha$ equivalent width quartile W$_{\mathrm {Ly\alpha}}=52.63$ with respect to their other three quartiles. However, they only find a decrease of $\sim30\%$ in C IV equivalent widths. Due to the increase in C II equivalent width with increasing Ly$\alpha$ equivalent width, this may indicate that the neutral gas has a larger covering fraction than the ionized gas, which likely has a larger range of velocity offsets. Our C IV and N V profiles also differ qualitatively from S03 and high-redshift UV-bright SFGs \citep{Heckman2011}, in that we find the profiles to be dominated by the interstellar component, as opposed to the stellar component.

As can be seen in the lower left panel of Figure \ref{Fig:Lya_dep}, for increasing Ly$\alpha$ equivalent width, the UV continuum slopes of the composite spectra become bluer. We do note that the spectral slopes of the VIMOS spectra were scaled so that their average spectral slope matched that of the FORS spectra. By fitting a power law to the $1250-1600{\mathrm \AA}$ region of the composite spectra, we find a UV spectral slope (F$_{\lambda}$ $\propto$ $\lambda^{\beta}$) of $\beta=-2.25$ and $\beta=-1.50$ for Ly$\alpha$ equivalent widths of W$_0=50.0{\mathrm \AA}$ and W$_0=8.4{\mathrm \AA}$. S03 also find a bluer continuum slope for galaxies with stronger Ly$\alpha$ emission. Similarly, we find a UV spectral slope of $\beta=-1.85$ and $\beta=-1.40$ for the highest and second highest Ly$\alpha$ equivalent width quartiles from S03 respectively. The bottom left panel of Figure \ref{Fig:Lya_dep} shows the relation between spectral slope index and Ly$\alpha$ equivalent width. Although we find the same trend of bluer continuum slopes with increasing Ly$\alpha$ equivalent width, we do find systematically bluer continuum slopes than S03, which may be impacted by the VIMOS flux calibration. Using photometrically determined $\beta$ values, we find a mean spectral slope among individual galaxies of $<\beta>=-1.52\pm0.45$ and $<\beta>=-1.22\pm0.47$ for the UV-bright LAE and non-LAE composites.

Across the UV-bright non-LAE and UV-bright LAE composites, we measure the average velocity offset between low-ionization absorption and Ly$\alpha$ emission to remain constant from $\Delta v_{{\mathrm {em-abs}}}=640\pm10$ km s$^{-1}$ to $\Delta v_{{\mathrm {em-abs}}}=610\pm60$ km s$^{-1}$ with increasing Ly$\alpha$ equivalent width. As is shown in the bottom right panel of Figure \ref{Fig:Lya_dep}, our measurements are plotted as red diamonds while those of S03 are in black. Our results are also consistent with S03, who found velocity offset to decrease monotonically with increasing Ly$\alpha$ equivalent width. We do note finding consistently higher velocity offsets than S03 at each Ly$\alpha$ equivalent width. \cite{Steidel2010} found a median velocity offset of $\Delta v_{{\mathrm {em-abs}}}=609$ km s$^{-1}$ with a scatter of 32km s$^{-1}$ for their sample of 89 z$\sim$2.3 SFGs. With the exception of baryonic and dynamical mass, they report not finding any significant correlations between velocity offset and other galaxy parameters.

        \subsection{Intrinsic UV Color Dependences}
\label{sec:UV_col_dep}

The spectral slope of the far-UV continuum is determined by the star formation rate history and the amount of extinction. A continuously star-forming galaxy has an unreddened UV spectral energy distribution shape that remains fairly constant for ages 10Myr to 1Gyr whose spectral slope, $\beta$, ranges from -2.6 to -2.1 \citep{Leitherer1999}. UV-bright, $2<z<3.5$, star-forming galaxies typically have ages within this range \citep{Papovich2001, Shapley2001, Erb2006, Gawiser2007, Kornei2010} indicating that large differences in UV spectral slope reflect varying amounts of dust extintion. Due to uncertainties in the star formation histories and the form of the dust extinction law, we report trends based on the spectral slope index. Continuum spectral slopes are too difficult to measure in individual spectra, however, broadband photometry allows us to parameterize our data set based on spectral slopes as calculated from intrinsic UV colors.

\begin{figure}[!t]
  \plotone{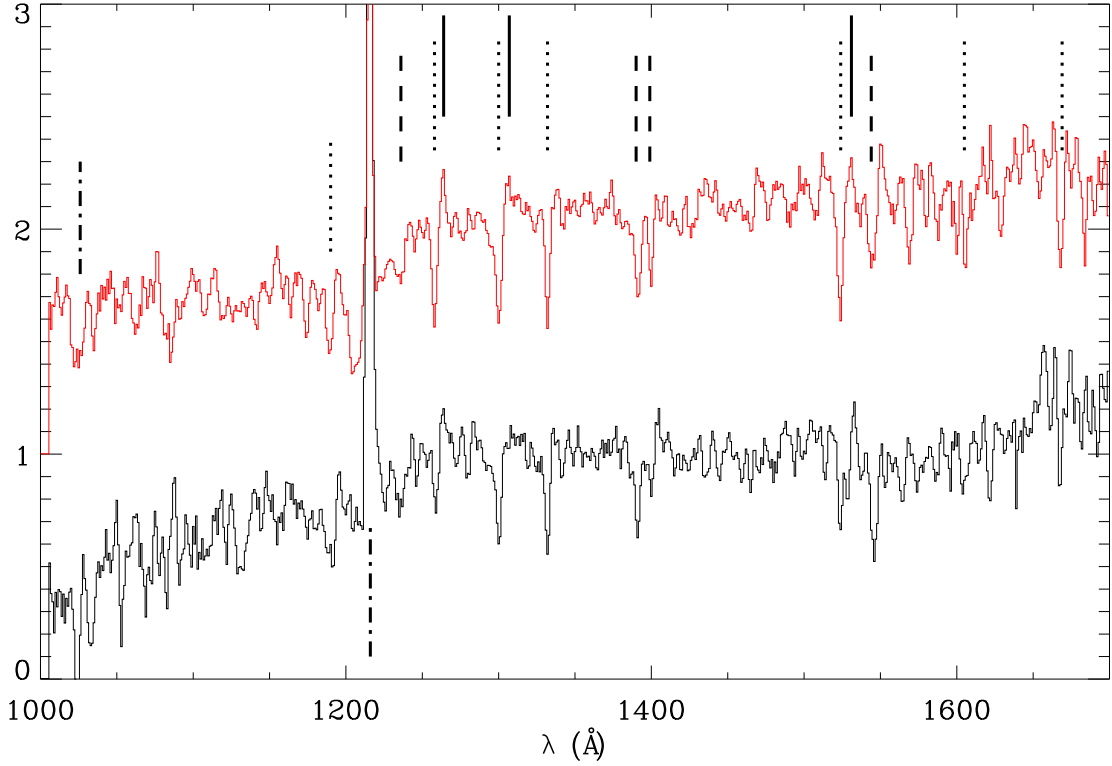}
    \caption{Composite spectra of SFGs divided into two subsamples based on intrinsic UV spectral slopes as calculated from photometric colors. Top - the red UV continuum slope composite ($\beta_{phot}>-1.4$). Bottom - the blue continuum slope composite ($\beta_{phot}\le-1.4$). With bluer spectral slope, Ly$\alpha$ emission increases, low-ionization absorption become weaker and there is significantly stronger Si IV $\lambda$1402 nebular emission.}
\label{Fig:B_SH_ebv_stack}
\end{figure}

\begin{figure}[!t]
  \plotone{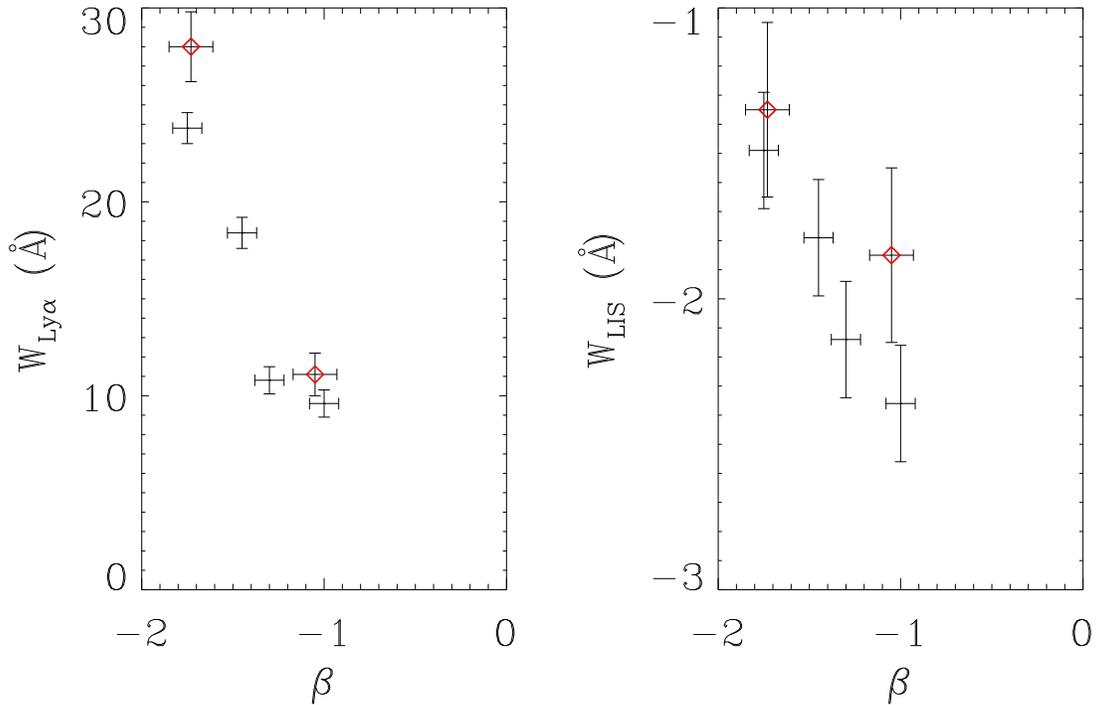}
    \caption{Left - the dependence of Ly$\alpha$ equivalent width on spectral slope, $\beta$, when the dataset is subdivided based on spectral slope for this work (red diamonds) and S03 (black points). Right - the dependence of low-ionization absorption on spectral slope.  The trends in decreasing Ly$\alpha$ emission and increasing low-ionization absorption for bluer spectral slopes observed by S03 are confirmed.}
\label{Fig:ebv_dep}
\end{figure}

Figure \ref{Fig:B_SH_ebv_stack} displays two composite spectra composed of the red ($\beta>-1.4$) and blue ($\beta \le -1.4$) half of our UV-bright galaxies. Through fitting a power law to the observed photometric colors, we find spectral slopes of $<\beta>=-1.05\pm0.34$ and $<\beta>=-1.73\pm0.34$ for the red and blue composites respectively where the errors represent the standard deviation. Since the spectral slopes of the VIMOS spectra were scaled so that their average spectral slope matched that of the FORS spectra, we use the photometric $\beta$ values when examining the relationship among galaxy properties and spectral slope. Figure \ref{Fig:ebv_dep} shows the dependence of Ly$\alpha$ equivalent width on $\beta$ (left panel) and the dependence of average low-ionization equivalent width on $\beta$ (right panel) for this work (red diamonds) and S03 (black points). With bluer spectral slope, Ly$\alpha$ equivalent width increases from W$_0= 11.0\pm1.1 {\mathrm \AA}$ to W$_0=27.0\pm1.8{\mathrm \AA}$. Similarly, the average low-ionization absorption line strength decreases from W$_{\mathrm {LIS}}= -1.83 {\mathrm \AA}$ to W$_{\mathrm {LIS}}= -1.35 {\mathrm \AA}$. We note that similar to the UV-bright LAE and non-LAE composites, while Si II absorption decreases significantly with bluer slopes, C II absorption does not. Additionally, the change in Ly$\alpha$ equivalent width is not sufficient to fully explain this decrease in low-ionization absorption strength, and therefore a correlation must exist between spectral slope and low-ionization absorption. In contrast to the UV-bright LAE composite, the ratio of Si II absorption line equivalent widths is W$_{0}(1260)$/W$_{0}(1526)=1.0\pm0.5$ and W$_{0}(1260)$/W$_{0}(1526)=0.8\pm0.2$ in the blue and red composites respectively, both consistent with unity. A ratio of unity indicates that Si II is optically thick, suggesting this is the case in both composites.

The Si IV $\lambda$1393 and C IV $\lambda$1549 high-ionization strengths do not vary substantially with spectral slope. With bluer spectral slope, the Si IV $\lambda$1402 absorption line strength decreases by $\sim$50\% due to P Cygni emission filling in the absorption. Moreover, relative to the red composite, the N V P Cygni profiles are more prominent. The N V $\lambda$ 1240 absorption equivalent width is W$_0=0.8\pm0.3$ with a large velocity offset, $\Delta v>1200$ km s$^{-1}$. Nebular features originate from H II regions at the systemic redshift and indicate a larger presence of high mass O- and B-type stars. On the other hand, we do not find a C IV $\lambda$1549 P Cygni profile in the blue spectral slope composite while we do observe one in the red one. Finally, there is no significant change in Si II$^{\ast}$ emission strength with spectral slope, in spite of a difference in Ly$\alpha$ equivalent width.

        \subsection{Redshift Evolution}
\label{sec:red_evol}

There are roughly the same number of UV-bright LAEs and UV-bright non-LAE objects at z$\sim$2 and z$\sim$3, allowing us to divide our sample based on redshift and Ly$\alpha$ emission, generating four composite spectra. The low-redshift sample (2$<z<$2.7) is composed almost entirely of FORS objects while the high-redshift sample (2.7$<z<$3.5) is mostly VIMOS objects. In Figure 10, the top panel shows the two UV-bright LAE composites for z$\sim$2 (red, upper) and z$\sim$3 (black, lower). The bottom panel is the same as the top panel except for the UV-bright non-LAE composites. In this figure, the Ly$\alpha$ dependences discussed in \S \ref{sec:Lya_dep} are also apparent. These trends are observed at both redshifts, as the two UV-bright LAE composites show significantly stronger low-ionization absorption. The S/N of the low-redshift composites degrades quickly blueward of Ly$\alpha$ due to poor atmospheric transmission and low throughput of the spectrograph below 3600${\mathrm \AA}$. We present equivalent widths, velocity offsets and continuum spectral slopes for these composites as well as a z$\sim$3 all-LBG and z$\sim$2 all-BX composites in Table \ref{Tab:z_prop}.

For UV-bright LAEs, there are several interesting trends with decreasing redshift, including weaker Ly$\alpha$ emission, stronger insterstellar absorption, stronger Si II$^{\ast}$ emission, and bluer spectral slopes. The change in Ly$\alpha$ emission line strength is the clearest trend, as from z$\sim$3 to z$\sim$2, the Ly$\alpha$ equivalent width decreases from W$_0=66.5\pm6.0 {\mathrm \AA}$ to W$_0=29.5\pm5.5 {\mathrm \AA}$. We find interpreting this result difficult as there are a multitude of effects that contribute to the strength of the Ly$\alpha$ emission line. However, \cite{Nilsson2009} and \cite{Guaita2010} both find LAEs at z$\sim$2 to appear more evolved than those at z$\sim$3, although they do find comparable photometric Ly$\alpha$ equivalent widths. Aside from Ly$\alpha$ emission, very few features are seen in the z$\sim$3 UV-bright LAE composite, which may be due to the faint continua of z$\sim$3 UV-bright LAEs resulting in a low S/N. For this reason, it is difficult to determine if the differences between these two composites reflects an evolution in galaxy properties. In the z$\sim$2 composite, the weak absorption strength of the Si II resonant lines are again apparent as the $\lambda$1260 transition is faint and there is no $\lambda$1526 absorption. 
In contrast, we do find evidence for Si II$^{\ast}$ fine-structure emission at $\lambda \lambda$ 1265, 1533, but again we do not see Si II$^{\ast}$ $\lambda$1309 emission. We also note that the relative strength of Si II$^{\ast}$ emission with respect to Si II resonant absorption is significantly higher in our low-redshift UV-bright LAE composite than either the high-redshift UV-bright LAE composite or the high-Ly$\alpha$ equivalent width quartile composite of S03, which was also composed of z$\sim$3 galaxies.

In the UV-bright non-LAE composites, we find little evolution in galaxy properties from z$\sim$3 to z$\sim$2 with the exception of stronger high-ionization absorption at higher redshifts. The high-ionization interstellar absorption strengths are on average $\sim50\%$ stronger in the high-redshift subsample. However, we note that the z$\sim$3 subsample is dominated by VIMOS LBGs with a larger instrument FWHM. As can be seen in Table \ref{Tab:z_prop}, we find comparable Ly$\alpha$ equivalent widths, low-ionization equivalent widths, and velocity offsets. Therefore, this discrepancy may not reflect an actual evolution in high-ionization strength. Furthermore, there is Si II$^{\ast}$ emission at $\lambda$1265 of the z$\sim$2 composite, although we find no other evidence for Si II$^{\ast}$ fine-structure emission at either redshift. Thus, the relative strength of Si II$^{\ast}$ emission for SFGs appears to be stronger at z$\sim$2 than z$\sim$3. As was previously seen in the UV-bright LAE and UV-bright non-LAE composites, Si II$^{\ast}$ emission is stronger with stronger Ly$\alpha$ emission.

As discussed in \S~\ref{sec:ind_gal_prop}, the spectral slopes of individual galaxies were measured based on their photometric colors for which we found bluer spectral slopes at lower redshift. This trend is marginally significant among the UV-bright non-LAE and LAE subsets. We find bluer mean spectral slopes of $\beta_{\mathrm LBG-only}=-0.97\pm0.58$ to $\beta_{\mathrm BX-only}=-1.39\pm0.24$ for UV-bright non-LAE galaxies and $\beta_{\mathrm LBG-LAE}=-1.34\pm0.33$ to $\beta_{\mathrm BX-LAE}=-1.87\pm0.51$ for UV-bright LAE galaxies (errors represent the standard deviation of the mean). The errors due to photometric uncertainties are comparable to the standard deviation of the mean. We do not attempt to measure the spectral slope from the continuum of the composite spectra as all of our high-redshift subsamples are composed almost entirely of VIMOS objects whose individual spectra were scaled to match the FORS spectra. Bluer spectral slopes at lower redshifts would indicate that UV-bright non-LAE galaxies and UV-bright LAEs are becoming less evolved or less dusty with cosmic time. 
While this trend of bluer spectral slopes with decreasing redshift is both surprising and interesting, more observations will be necessary to understand if it is real. Finally at each redshift, we also find bluer spectral slopes with stronger Ly$\alpha$ emission indicating a connection between line and continuum extinction at both redshifts.

\section{Conclusions}
\label{sec:conclusions}

We have combined and analyzed the spectra of both narrowband- and continuum-selected star-forming galaxies across a redshift range of $2<z<3.5$. Through placing a canonical flux limit of R$<$25.5, we are able to study the relationship among Ly$\alpha$ emission, intrinsic UV spectral slope, redshift, outflow kinematics, low- and high-ionization absorption, and nebular emission for a sample of galaxies spanning a similar luminosity range. Our results are summarized below.

\begin{itemize}
\item UV-bright SFGs have spectral slope indices, $\beta$, that span a large range from $-2.7$ to $0.1$ while UV-bright LAEs (W$_{\mathrm {Ly\alpha}}>20{\mathrm \AA}$) have bluer spectral slope indices with $-2.7<\beta<-0.8$, in good agreement with model predictions for a young dust-free SFG \citep[$\beta \sim -2.6$,][]{Leitherer1999}. Of the 59 R$<$25.5 SFGs, 10 have Si II$^{\ast}$ emission and 12 have multiple-peaked Ly$\alpha$ emission. Additionally, high-ionization absorption lines are dominated by the interstellar, as opposed to the stellar component.

\item UV-bright LAEs with rest-frame W$_{\mathrm {Ly\alpha}}>20{\mathrm \AA}$ have weaker low-ionization absorption, weaker C IV absorption, bluer spectral slopes, and stronger Si II$^{\ast}$ fine-structure emission than UV-bright non-LAE galaxies with rest-frame W$_{\mathrm {Ly\alpha}}<20{\mathrm \AA}$. Among low- and high-ionization absorption strengths, individual elements also show different dependences on Ly$\alpha$ emission.

\item SFGs with bluer spectral slopes have more prominent N V $\lambda$$\lambda$ 1238,1242 and Si IV $\lambda$1402 P Cygni profiles, weaker low-ionization absorption, and stronger Ly$\alpha$ emission.

\item From $2.7<z<3.5$ to $2.0<z<2.7$, SFGs exhibit bluer spectral slopes and stronger Si II$^{\ast}$ fine-structure emission. Additionally, those with rest-frame W$_{\mathrm {Ly\alpha}}<20{\mathrm \AA}$ have comparable low- and high-ionization absorption strengths as well as similar Ly$\alpha$ emission strengths. In contrast, there is a significant decrease in Ly$\alpha$ emission strength in the subsample with rest-frame W$_{\mathrm {Ly\alpha}}>20{\mathrm \AA}$, but we caution this may be due to selection effects. 
\end{itemize}

The observed trends with Ly$\alpha$ emission are largely consistent with the physical picture put forth by S03 where the mechanical energy from star formation and supernovae in high-redshift UV-bright SFGs is powering galaxy-scale outflows of ionized and neutral gas. Evidence for these outflows is seen as strong blueshifted absorption as well as redshifted H I Ly$\alpha$ emission relative to nebular features at the systemic redshift. Amongst the individual galaxies, a wide range of velocity offsets between Ly$\alpha$ emission and interstellar absorption from $\Delta v_{\mathrm {em-abs}}\sim200$ km s$^{-1}$ to $\Delta v_{\mathrm {em-abs}}\sim900$ km s$^{-1}$ is observed. In spite of this range in velocity offsets, we do not find a significant change in $\Delta v_{\mathrm {em-abs}}$ with respect to Ly$\alpha$ emission strength in the composite spectra. We note that our velocity offsets decrease slightly with Ly$\alpha$ emission and are also consistent with S03 who found velocity offset to decrease monotonically with increasing Ly$\alpha$ emission strength. The velocity offset of our UV-bright SFG composite of $\Delta v_{\mathrm {em-abs}}=630$ km s$^{-1}$ is also remarkably similar to the average velocity offset of $\Delta v_{\mathrm {em-abs}}=609$ km s$^{-1}$ from \cite{Steidel2010} for a sample of 89 continuum-selected galaxies at $z\sim2.3$.

Within the outflowing component, the high-ionization features have similar kinematic properties to and span a similar velocity range as the low-ionization features. However, this information says nothing about the physical distribution of the absorbing gas. \cite{Steidel2010} find significant absorption of both low- and high-ionization species displaying similar radial dependences at galactocentric radii spanning $3-125$kpc. This indicates that both ionized and neutral gas share not only the same velocity envelope, but also similar physical distributions.

Using the low-ionization absorption lines as a probe of the outflowing neutral gas, we find a direct correlation between Ly$\alpha$ emission strength and the average strength of low-ionization absorption. This trend is most clearly demonstrated by Si II ions. In the outflowing component, a difference in line strengths indicates a change in either the velocity dispersion, covering fraction, or both. While the Si IV transition is consistent with being optically thin in all composites, the Si II transition and the other low-ionization transitions are optically thick. We do note that among galaxies with rest-frame W$_{\mathrm {Ly\alpha}}>20{\mathrm \AA}$, Si II absorption is significantly weaker and the $\lambda$1526 transition in the UV-bright LAE composite is marginally detected leading to a Si II $\lambda$1260 to $\lambda$1526 ratio that is greater than one, which would be expected if Si II were not optically thick. While \cite{Steidel2010} found Si II to be unsaturated at $b\sim63$kpc, those authors did find it to be saturated at $b\sim3$kpc and $b\sim40$kpc. We also note that due to the low S/N, the ratio of these two line strengths is fairly uncertain. Given the similar kinematic properties of different Si ionization states, different ionization state dependences on Ly$\alpha$ emission, and similar physical distributions, we find it likely that the outflowing neutral gas is in the form of neutral clouds embedded in ionized gas as previously theorized by \cite{Steidel2010}.

Similar to the results of S03, we find a significant correlation between bluer spectral slopes and weaker low-ionization absorption strength. SFGs with ages between 10Myr and 1Gyr and constant star-formation histories have similar UV spectral slopes, so the shape of the observed spectral slope is largely dependent on dust extinction. UV-bright SFGs and LAEs at z$\sim$2 and z$\sim$3 are well-characterized by these ages and star formation histories \citep{Shapley2001, Lai2008, Nilsson2009, Guaita2010}. Therefore, we interpret this result as evidence for the presence of dust in the outflowing component. In addition, these outflows must also cover a region sufficiently large to extinguish a significant amount of the UV continuum surface brightness. The increased prominence of nebular features for bluer $\beta$ values is not surprising as bluer spectral slopes indicate either a lower amount of extinction, the presence of hotter more massive stars, or both. Each of these effects would make stellar photospheric features more visible.

From a redshift of $2.7<z<3.5$ to $2.0<z<2.7$, the UV-bright non-LAE spectra look remarkably similar. They have comparable Ly$\alpha$ emission strengths, interstellar absorption strengths, and kinematics consistent with no significant evolution in SFGs with rest-frame W$_{\mathrm {Ly\alpha}}<20{\mathrm \AA}$ or the nature of their outflows across this redshift interval. In contrast, for SFGs with W$_{\mathrm {Ly\alpha}}>20{\mathrm \AA}$ we find significantly weaker Ly$\alpha$ emission in the lower redshift sample. If this trend is real, then it is consistent with UV-bright LAEs being more evolved at lower redshifts. Low- and high-ionization absorption also appear to be stronger in the low-redshift sample, although due to the faint continua of z$\sim$3 UV-bright LAEs, this may not indicate an evolution in galaxy properties. For both subsets of SFGs, the lower-redshift composites have bluer spectral slopes, which is confirmed at marginal significance for the UV-bright LAE composite from the photometric spectral slopes of individual galaxies. As there are many systematic effects that can contribute to the spectral slope, additional investigation specifically a better constraint on their rest-UV spectral slopes would be needed to conclude that lower redshift continuum- and narrowband-selected galaxies are less evolved, as their bluer colors would imply.

We thank Alice Shapley and Kim Nilsson for graciously sharing their data. 
We would also like to extend special thanks to Peter Kurczynski, Viviana Acquaviva, and Jean Walker-Soler for helpful discussions and suggestions that improved the quality of this paper.
This work was supported by NSF grants AST 08-07570 and AST 08-07885.
The Institute for Gravitation and the Cosmos is supported by the Eberly College of Science and the Office of the Senior Vice President for Research at the Pennsylvania State University.

\bibliographystyle{apj}   
\bibliography{apj-jour,Berry11_LAEspec}


\begin{table}[h]
\caption{Subsample Criteria}
\centering
\begin{tabular}{l c c c c c}
\hline \hline
Type   & Magnitude & W$_{\mathrm {Ly\alpha}}$ & N$_{\mathrm {FORS}}$ & N$_{\mathrm {VIMOS}}$ &N$_{\mathrm {total}}$\\
\hline

UV-bright SFG  & $R\le 25.5$ & all                    & 29 & 30 & 59 \\
UV-bright non-LAE & $R\le 25.5$ & $<20 {\mathrm \AA}$    & 15 & 12 & 27 \\
UV-bright LAE  & $R\le 25.5$ & $\ge 20 {\mathrm \AA}$ & 14 & 18 & 32 \\
UV-bright blue & $R\le 25.5$ & -                      & 14 & 11 & 25 \\
UV-bright red  & $R\le 25.5$ & -                      & 11 & 17 & 28 \\
UV-faint LAE   & $R>25.5$    & $\ge 20 {\mathrm \AA}$ & 14 & 8  & 22 \\

\hline
\end{tabular}
\label{Tab:sel_crit}
\end{table}

\begin{table}[h]
\caption{All-SFG Spectroscopic Features}
\centering
\begin{tabular}{l c c c c c}
\hline \hline
Ion & $\lambda_{\mathrm {rest}}$ (${\mathrm \AA}$) & $f$ & $W_{0}$ (${\mathrm \AA}$) & $\sigma$ (${\mathrm \AA}$) & $\Delta v_{\mathrm {em-abs}}$ (km s$^{-1}$) \\

\hline
Ly$\alpha$$_{\mathrm {em}}$  & 1215.67 & - & 19.1 & 1.5 & -  \\
Ly$\alpha$$_{\mathrm {abs}}$ & 1215.67 & - & -2.7 & 0.3 & 2400  \\
Si II  &  1260.42 & 1.007  & 1.3 & 0.15 & 520   \\
O I    &  1302.17 & 0.0489 & 2.1 & 0.2 & 760   \\
Si II  &  1304.37 & 0.094  & 2.1 & 0.2 & 760   \\
C II   &  1334.53 & 0.128  & 1.7 & 0.2 & 590   \\
Si II  &  1526.71 & 0.130  & 1.2 & 0.3 & 580   \\
Fe II  &  1608.45 & 0.058  & 0.7 & 0.3 & 590   \\
Al II  &  1670.79 & 1.83   & 1.2 & 0.4 & 560   \\
N V    &  1238.82 & 0.157  & 0.9 & 0.2 & 1320   \\
N V    &  1242.80 & 0.0782 & 0.9 & 0.2 & 1320   \\
Si IV  &  1393.76 & 0.514  & 1.5 & 0.2 & 550   \\
Si IV  &  1402.77 & 0.255  & 0.9 & 0.2 & 630   \\
C IV   &  1548.20 & 0.191  & 1.7 & 0.3 & 800   \\
C IV   &  1550.78 & 0.0952 & 1.7 & 0.3 & 800   \\

\hline
\end{tabular}
\tablecomments{Ly$\alpha$$_{{\mathrm r}}$ and Ly$\alpha$$_{{\mathrm b}}$ refer to the red and blue components of the Ly$\alpha$ profile. All values of $\Delta v$ are relative to Ly$\alpha$$_{{\mathrm r}}$. Transition oscillator strengths as in \cite{Pettini2002}. $W_{0}$, $\sigma$, and $\Delta v$ values listed for O I $\lambda$1302 and Si II $\lambda$1304 refer to the line profile for these two blended features assuming the rest wavelength is $\lambda=1303.27{\mathrm \AA}$. Similarly, the values for N V $\lambda$$\lambda$ 1238.82, 1242.80 and C IV $\lambda$$\lambda$ 1548.20, 1550.78 refer to the blended line profiles assuming a rest wavelength of $\lambda=1240.81{\mathrm \AA}$ and $\lambda=1549.48{\mathrm \AA}$ respectively.}
\label{Tab:all_SFG_prop}
\end{table}

\begin{table}[h]
\caption{Spectroscopic Properties}
\centering
\begin{tabular}{l c c c}
\hline \hline
   & UV-bright non-LAE & UV-bright LAE & UV-bright SFG\\
\hline
$N_{gal}$ & 27 & 32 & 59\\
$R$ & 24.46 & 24.84 & 24.72  \\
$\beta_{phot}$ & -1.3 & -1.5 & -1.3  \\
$\beta_{spec}$ & -1.50 & -2.25 & -1.90  \\
$\Delta v_{{\mathrm {em-abs}}}$ & $640\pm10$ & $610\pm60$ & $630\pm20$  \\
FWHM$_{LIS}$ & $340\pm100$ & $420\pm90$ & $430\pm60$  \\

$W_{{\mathrm {Ly\alpha_r}}}$ & $8.4\pm0.9$ & $50.0\pm4.8$ & $19.1\pm1.5$  \\
$W_{{\mathrm {Ly\alpha_b}}}$ & $-4.9\pm0.9$ & $1.2\pm0.8$ & $-2.7\pm0.3$  \\
$W_{{\mathrm {SiII,1193}}}$ & $-1.9\pm0.3$ & -  & $2.0\pm0.2$\\
$W_{{\mathrm {SiII,1260}}}$ & $-1.4\pm0.25$ & $-1.2\pm0.2$ & $-1.3\pm0.15$  \\
$W_{{\mathrm {OI+SiII,1303}}}$ & $-2.5\pm0.3$ & $-1.7\pm0.3$ & $-2.1\pm0.2$  \\
$W_{{\mathrm {CII,1334}}}$ & $-1.5\pm0.25$ & $-1.8\pm0.3$ & $1.7\pm0.2$  \\
$W_{{\mathrm {SiII,1526}}}$ & $-1.8\pm0.3$ & $-0.5\pm0.3$ & $1.2\pm0.2$  \\
$W_{{\mathrm {FeII,1608}}}$ & $-1.1\pm0.4$ & $-$          & $0.7\pm0.3$  \\
$W_{{\mathrm {AlII,1670}}}$ & $-1.5\pm0.45$ & $-1.0\pm0.5$ & $1.2\pm0.3$  \\
$W_{{\mathrm {NV,1240}}}$ & $-1.0\pm0.35$ & $-1.1\pm0.4$ & $0.9\pm0.2$  \\
$W_{{\mathrm {SiIV,1393}}}$ & $-1.5\pm0.3$ & $-1.4\pm0.35$ & $1.5\pm0.2$  \\
$W_{{\mathrm {SiIV,1402}}}$ & $-1.0\pm0.3$ & $-0.4\pm0.35$ & $0.9\pm0.2$  \\
$W_{{\mathrm {CIV,1549}}}$ & $-2.2\pm0.35$ & $-1.1\pm0.4$ & $1.7\pm0.3$ \\

$W_{{\mathrm {OIII,1664}}}$ & - & $ 0.8\pm0.45 $ & -  \\
\hline
\end{tabular}
\tablecomments{R and $\beta_{phot}$ refer to the median values of the individual galaxies. $\beta_{spec}$ refers to the spectral slope of the composite. The UV-bright LAE and UV-bright non-LAE subsets contain a similar fraction of galaxies observed with FORS and VIMOS. $\Delta v_{{\mathrm {em-abs}}}$ and FWHM$_{LIS}$ refer to the average values of the low-ionization lines.}
\label{Tab:comp3_prop}
\end{table}

\begin{table}[h]
\caption{Si II$^{\ast}$ Emission in UV Bright SFG Composite}
\centering
\begin{tabular}{l c c c}
\hline \hline
 $\lambda_{rest}$   & $A_{ul}$            & $W_{0}$           & $\Delta v$   \\
 (${\mathrm \AA}$) & ($10^{8}$ s$^{-1}$) & (${\mathrm \AA}$) & (km s$^{-1}$) \\
\hline
1264.74 & 23.0 & 0.6$\pm$0.2 & 280 \\
1309.28 & 7.00 & 0.0$\pm$0.2 &  -   \\
1533.43 & 7.40 & 0.7$\pm$0.3 & 140 \\
\hline
\end{tabular}
\tablecomments{Properties of Si II$^{\ast}$ Emission in the UV-bright SFG composite. Eistein $A-$coefficients from the NIST Atomic Specta Database (http://physics.nist.gov/cgi-bin/AtData/main\_asd). The Si II$^{\ast}$ $\lambda$1309 feature is not seen, so we do not report a $\Delta v$.}
\label{Tab:SiII}
\end{table}

\begin{sidewaystable}
\caption{Spectroscopic Properties of z$\sim$2 and z$\sim$3 UV Bright Composites}
\centering
\begin{tabular}{l c c c c c c}
\hline \hline
   & z$\sim$2 non-LAE & z$\sim$3 non-LAE & z$\sim$2 LAE & z$\sim$3 LAE & z$\sim$2 SFG & z$\sim$3 SFG \\
\hline
$N_{gal}$ & 16 & 11 & 12 & 20 & 28 & 31 \\
$R$     & 24.55 & 24.36 & 24.69 & 24.94 & 24.62 & 24.85  \\
$\beta_{phot}$ & -1.3 & -1.2 & -1.7 & -1.4 & -1.4 & -1.2 \\
$\Delta v_{{\mathrm {em-abs}}}$ & 640 & 740 & 600 & 590 & 630 & 660  \\

$W_{{\mathrm {Ly\alpha_r}}}$ & $8.2\pm0.7$   & $10.6\pm1.4$  & $29.5\pm 5.5$ & $66.5\pm6$ & $12.3\pm0.8$ & $52.3\pm3.5$ \\
$W_{{\mathrm {Ly\alpha_b}}}$ & $-4.1\pm0.5$  & $-5.0\pm1.5$  & $1.8\pm 1.3$  &      -     & $-3.9\pm0.9$ & -           \\
$W_{{\mathrm {SiII,1193}}}$  & $-2.2\pm0.4$  & $-1.6\pm0.6$  & $-$          &    $-$      & $-2.8\pm0.4$ & -           \\
$W_{{\mathrm {SiII,1260}}}$  & $-1.3\pm0.2$  & $-2.0\pm0.25$ & $-1.2\pm0.4$ & -           & $-1.3\pm0.15$ & $-1.2\pm0.25$ \\
$W_{{\mathrm {OI+SiII,1303}}}$ & $-2.6\pm0.3$ & $-2.0\pm0.3$ & $-2.2\pm0.4$ & $-0.8\pm0.5$ & $-2.3\pm0.2$ & $-1.1\pm0.3$ \\
$W_{{\mathrm {CII,1334}}}$   & $-1.5\pm0.25$ & $-1.6\pm0.25$ & $-2.3\pm0.5$ & $-$         & $-1.7\pm0.2$ & $-1.3\pm0.3$ \\
$W_{{\mathrm {SiII,1526}}}$  & $-1.7\pm0.4$  & $-2.0\pm0.4$  & $-0.6\pm0.5$ & $-$         & $-1.3\pm0.25$ & $-1.0\pm0.3$ \\
$W_{{\mathrm {NV,1240}}}$    & $-1.1\pm0.5$  & $-$           & $-1.0\pm0.3$ & $-0.8\pm0.4$ & $-1.0\pm0.25$ & -          \\
$W_{{\mathrm {SiIV,1393}}}$  & $-1.4\pm0.35$ & $-1.5\pm0.45$ & $-1.6\pm0.5$ & $-$         & $-1.7\pm0.25$ & $-1.5\pm0.35$ \\
$W_{{\mathrm {SiIV,1402}}}$  & $-0.9\pm0.35$ & $-1.5\pm0.45$ & $-0.4\pm0.5$ & $-$         & $-0.9\pm0.25$ & $-0.7\pm0.35$ \\
$W_{{\mathrm {CIV,1549}}}$   & $-2.1\pm0.4$  & $-3.4\pm0.45$ & $-1.5\pm0.6$ & $-0.9\pm0.5$ & $-1.8\pm0.3$ & $-2.0\pm0.3$ \\
\hline
\end{tabular}
\caption{Analogous to Table~\ref{Tab:comp3_prop}. We do not report equivalent widths for features not seen in the composite spectra.}
\label{Tab:z_prop}
\end{sidewaystable}


%

\begin{figure}[!t]
  \plotone{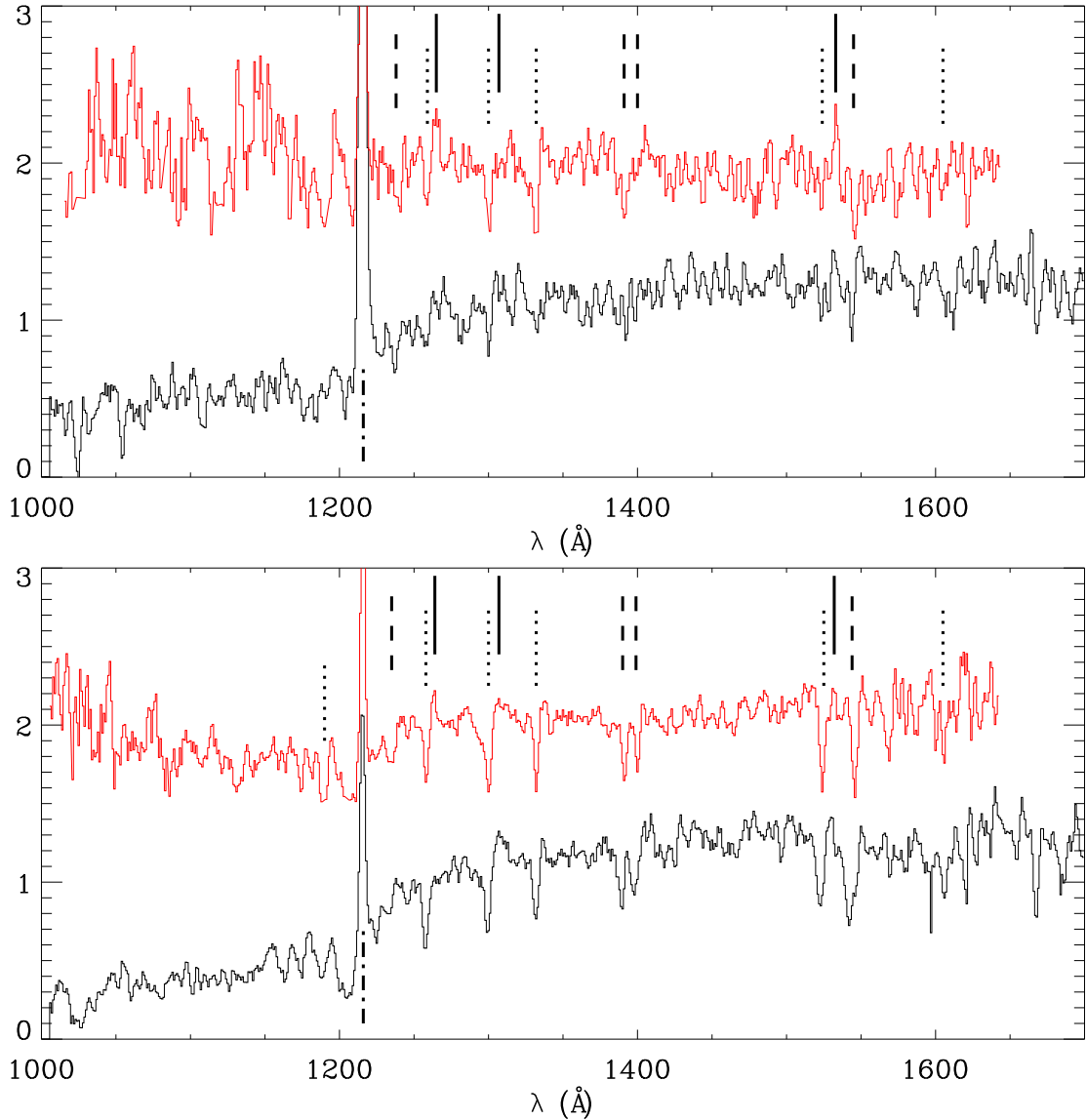}
    \caption{Composite spectra of UV-bright LAEs and UV-bright non-LAE objects for $2.0<z<2.7$ BXs in red and $2.7<z<3.5$ LBGs in black. Top - a composite of 13 BX-LAEs is plotted in red above a composite of 21 LBG-LAEs in black. For our sample of galaxies, few features are seen in the z$\sim$3 composite, however we do see a decrease in Ly$\alpha$ emission with redshift. In spite of the low S/N of the BX-LAE composite, the relative strength of Si II$^{\ast}$ $\lambda \lambda$ 1265, 1533 emission to resonant Si II absorption is significantly higher than any other composite. Bottom - the BX-only composite composed of 15 BXs is plotted in red above the LBG-only composite consisting of 11 LBGs in black. These composite spectra are quite similar and show no significant evolution of spectroscopic properties across this redshift interval.}
\label{Fig:B_SH_z_stack}
\end{figure}

\end{document}